\documentclass
[prc,twocolumn,superscriptaddress,showpacs,twoside,floatfix]{revtex4-1}
\usepackage{graphicx}
\usepackage{amsmath}

\begin{document}
\title{Two-proton radioactivity and three-body decay. V. Improved momentum
 distributions.}

\author{L.\ V.\ Grigorenko}
\affiliation{Flerov Laboratory of Nuclear Reactions, JINR, RU-141980 Dubna,
Russia}
\affiliation{Gesellschaft f\"{u}r Schwerionenforschung mbH, Planckstrasse 1,
D-64291, Darmstadt, Germany}
\affiliation{RRC ``The Kurchatov Institute'', Kurchatov sq.\ 1, 123182 Moscow,
Russia}

\author{I.\ A.\ Egorova}
\affiliation{Bogolubov Laboratory of Theoretical Physics, JINR, RU-141980 Dubna,
Russia}

\author{M.\ V.\ Zhukov}
\affiliation{Fundamental Physics, Chalmers University of Technology, S-41296
G\"{o}teborg, Sweden}

\author{R.\ J.\ Charity}
\affiliation{Department of Chemistry, Washington University, St.\ Louis,
Missouri 63130, USA}

\author{K.\ Miernik}
\affiliation{Faculty of Physics, University of Warsaw, 00-681 Warsaw, Poland}

\date{\today. \texttt{File: /coul3/coul3-ex/resubmit/coul3-ex-14-resubmit.tex }}

\begin{abstract}
Nowadays quantum-mechanical theory allows one to reliably calculate the 
processes of $2p$ radioactivity (true three-body decays) and the corresponding 
energy and angular correlations up to distances of the order of $10^3$ fm. 
However, the precision of modern experiments has now become sufficient to 
indicate some deficiency of the predicted theoretical distributions. In this 
paper we discuss the extrapolation along the classical trajectories as a method 
to improve the convergence of the theoretical energy and angular correlations at 
very large distances (of the order of atomic distances), where only the 
long-range Coulomb forces are still operating. The precision of this approach is 
demonstrated using the ``exactly'' solvable semianalytical models with 
simplified three-body Hamiltonians. It is also demonstrated that for heavy $2p$ 
emitters, the $2p$ decay momentum distributions can be sensitive to the effect 
of the screening by atomic electrons. We compare theoretical results with 
available experimental data.
\end{abstract}


\maketitle


\section{Introduction}


Two-proton radioactivity is the most recently discovered radioactive decay mode 
of nuclei and it is a very actively developing field. There were 42 years 
between the prediction \cite{gol60} and discovery \cite{pfu02,gio02} of $2p$ 
radioactivity and, subsequentially, seven years latter we have several well 
studied examples. A number of experiments performed in the last 2-3 years can be 
characterized as key for the field. In particular, correlations in the $2p$ 
decays have been  measured recently in $^6$Be \cite{gri09d}, $^{16}$Ne 
\cite{muk08}, $^{19}$Mg \cite{muk07,muk08}, $^{45}$Fe \cite{mie07}, and 
$^{94}$Ag \cite{muk06} providing qualitatively new information about the $2p$ 
decays. With correlation information becoming available, the $2p$ decay studies 
are now turning into a field of research where precise information about 
structure and continuum dynamics can be obtained. It is clear that our ability 
to extract useful information from correlations is directly dependent on how 
well we understand the propagation of particles in the long-range three-body 
Coulomb field.

\begin{figure}[tbp]
\includegraphics[width=0.48\textwidth]{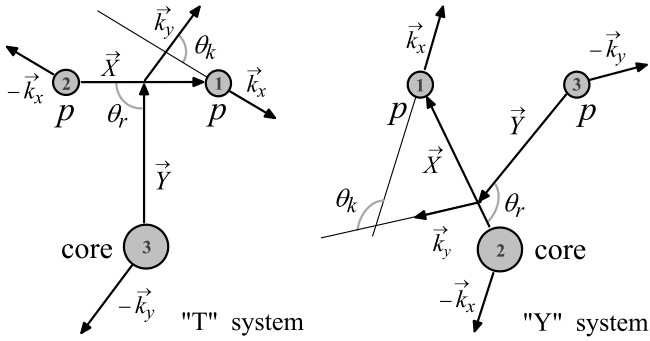}
\caption{Independent ``T'' and ``Y'' Jacobi systems for the core+$N$+$N$
three-body system in coordinate and momentum spaces. There are ``planar'' cases 
where both the coordinates and the momenta belong to the same plane.}
\label{fig:Jacobi}
\end{figure}

From a theoretical point of view, true two-proton decay ($2p$ radioactivity) is 
an exclusively quantum-mechanical phenomenon, which has no analogue in classical 
physics. It is expected to be widely spread along the proton drip line  with $Z 
< 50$ due to peculiarities of the pairing interaction. A consistent 
quantum-mechanical theory of two-proton radioactivity and ``democratic'' 
three-body decays of the coulombic nuclear systems has been developed in the 
series of papers \cite{gri00b,gri01a,gri03c,gri07,gri07a}, which we continue 
here, and has been applied to different physical cases in Refs.\ 
\cite{gri02,gri03,gri03a,gri09,gri09d}. The complete momentum correlations for 
the decay of a non-aligned three-body system can be described by two parameters. 
These parameters are chosen in this and our previous studies as the energy 
distribution parameter $\varepsilon $ between any two of the particles and the 
angle $\theta _{k}$ between the Jacobi momenta:
\begin{eqnarray}
\varepsilon = E_x/E_T \quad ,\quad \cos(\theta_k)=(\mathbf{k}_{x} \cdot
\mathbf{k}_{y}) /(k_x\,k_y) \, , \qquad \\
\label{eq:corel-param}
E_T =E_x+E_y=k^2_x/2M_x + k^2_y/2M_y \, , \nonumber  \\
M_x=\frac{A_1 A_2}{A_1+A_2}\,M \quad ,\quad
M_y=\frac{(A_1+A_2) A_3} {A_1+A_2+A_3}\,M \, , \nonumber  \\
{\bf k}_x  =  \frac{A_2 {\bf k}_1-A_1 {\bf k}_2 }{A_1+A_2} \, ,  \,\;
{\bf k}_y  =  \frac{A_3 ({\bf k}_1+{\bf k}_2)-(A_1+A_2) {\bf k}_3}
{A_1+A_2+A_3},\nonumber
\end{eqnarray}
where $A_i$ are mass numbers of the constituents, $M$ is a nucleon mass, and 
$E_T \equiv Q_{2p}$ is a two-proton decay energy. For two-proton emitters these 
parameters can be constructed in two ``irreducible'' Jacobi systems, called 
``T'' and ``Y'', see Fig.\ \ref{fig:Jacobi}. The detailed definition of the 
Jacobi coordinates can be found in  Ref.\ \cite{gri09d}. The complete 
correlation pictures for two-proton decay were, for the first time, calculated 
in Ref.\ \cite{gri03c}. Various aspects of the correlations between the decay 
products have been discussed in the theoretical works of 
Refs.~\cite{gri02,muk08,gri03c,gri07,gri09d}.

In $^6$Be and $^{45}$Fe, the complete correlation pictures for $2p$ decay were 
recently obtained experimentally \cite{gri09d,mie07}. Moreover, the precision of 
these experimental results is now sufficient to show a deficiency in certain 
aspects of the predicted momentum distributions in the case of heavy $2p$ 
emitters \cite{gri09}. It was already understood in Ref.\ \cite{gri01a} that 
this deficiency is connected to the limited radial range of the calculations and 
the approximate nature of the boundary conditions employed for the treatment of 
the three-body Coulomb asymptotic.

The classical extrapolation (CE) of momentum distributions was suggested in
Ref.\ \cite{gri03c} as a simple way to estimate the possible influence of the
``residual'' Coulomb interaction. The basic idea is that, at small distances,
particles are propagated by quantum-mechanical equations providing the
three-body wave function (WF) $\Psi_3^{(+)}$ with outgoing asymptotic. At some
sufficiently large distance, the WFs are converted into ``events'' with definite
coordinates and momenta by a Monte-Carlo (MC) procedure. However, at that time
(in 2002 the $2p$ decay of $^{45}$Fe was just discovered with statistics of the
order of ten events \cite{pfu02,gio02}) the need to improve this aspect of our
calculations was assigned to the remote future and no detailed studies were
performed. Now it seems that the development of the field has achieved the stage
where the need to improve this aspect of our approach has become evident.

In this work we discuss the method of classical extrapolation in detail, 
demonstrate its reliability by application to exactly-solvable three-body models 
with a simplified Hamiltonian, and consider three ``key'' cases ($^{6}$Be, 
$^{19}$Mg, and $^{45}$Fe) covering a broad range of possible charges, masses, 
and structures for the  $2p$ emitters.

Natural system of units with $\hbar=c=1$ is used in this work.


\section{Approximate boundary conditions}
\label{sec:approx}


In this section we sketch the methods  used to construct the approximate
boundary conditions \cite{gri01a} and outline existing problems. The asymptotic
form of the three-body potentials in the hyperspherical harmonics (HH) method is
\begin{eqnarray}
V_{K \gamma, K' \gamma'}(\rho) = \frac{U_{K \gamma, K' \gamma'}}{\rho^{3+N_{K
\gamma, K' \gamma'}}} + \frac{\mathcal{L}(\mathcal{L}+1)}{\rho^2}\delta_{K
\gamma, K' \gamma'} \nonumber \\
+ \frac{v \eta_{K \gamma, K' \gamma'}}{\rho} \,,
\label{eq:pot-ass}
\end{eqnarray}
where multiindex $\{K \gamma\}=\{K,L,S,l_x, l_y,s_x\} $ is a complete set of 
quantum numbers. The matrix $U_{K \gamma, K' \gamma'}$ arises due to 
contributions from the short-range nuclear forces, and $N_{K \gamma, K' \gamma'} 
\geq 0$ are some integer numbers. The effective contribution of the short-range 
forces decreases as $\rho^{-3}$ or faster in hypersherical space. The diagonal 
centrifugal term depends on the ``effective angular momentum'' 
$\mathcal{L}=K+3/2$. Coulomb pairwise potentials generate the long-range part of 
the hyperspherical potentials behaving as $\rho^{-1}$. From the technical side, 
the three-body Coulomb interaction causes problems due to long-range channel 
coupling (nonzero nondiagonal ``Sommerfeld parameters'' $\eta_{K \gamma, K' 
\gamma'}$) that does not allow one to decouple the HH equations on the 
asymptotic. To deal with this problem, the finite-size potential matrix (in 
truncated hyperspherical basis) can be diagonalized with respect to the 
long-range term by the orthogonal transform $\tilde{V}=A^T V A$:
\begin{equation}
\tilde{V}_{K \gamma, K' \gamma'}(\rho) = \frac{\tilde{U}_{K \gamma, K'
\gamma'}}{\rho^{3}} + \frac{C_{K \gamma, K' \gamma'}}{\rho^2} + \frac{v \eta_{K
\gamma}}{\rho}\delta_{K \gamma, K' \gamma'} \,.
\label{eq:pot-ass-diag}
\end{equation}
This potential includes nondiagonal ``centrifugal'' terms $C_{K \gamma, K'
\gamma'}$ and, to achieve the asymptotic in the diagonalized representation, we
still need to go very far in $\rho$ value, where the terms  $\sim \rho^{-2}$
become negligible compared to those with $\sim \rho^{-1}$. At such $\rho$ 
values, the hyperradial part of the asymptotic solution with pure outgoing 
nature can be constructed in the form
\begin{eqnarray}
\chi^{(+)}_{ K \gamma}(\rho)\sim\sum_{ K' \gamma'}A_{K \gamma, K' \gamma'}\left(
G_{\mathcal{L}_0}(\eta_{K' \gamma'},\rho)+iF_{\mathcal{L}_0}(\eta_{K'
\gamma'},\rho) \right), \nonumber \\
\Psi_3^{(+)}= \rho^{-5/2} \sum_{ K \gamma} \chi^{(+)}_{ K \gamma}(\rho)
\mathcal{J}_{ K \gamma}(\Omega_5) \,.\qquad
\label{eq:wf-ass}
\end{eqnarray}
The functions $F$ and $G$ are the ordinary regular and irregular Coulomb 
functions. Hyperspherical harmonics $\mathcal{J}_{ K \gamma}$ are functions of 
the 5-dimensional ``solid angle'' $\Omega_5=\{\theta_{\rho},\Omega_x, 
\Omega_y\}$. Here $\Omega_x$ and $\Omega_y$ are ordinary solid angles of the 
Jacobi vectors $\mathbf{X}$ and $\mathbf{Y}$ [see Eq.\ \ref{eq:hh-coord}] and 
$\tan(\theta_{\rho})=\sqrt{M_x/M_y} X/Y$. The value $\mathcal{L}_0$ should be 
larger than 3/2, but otherwise does not seem to be particularly important. The 
WFs $\chi^{(+)}$ provide the necessary boundary conditions for the decay 
problem.

The proposed boundary conditions are exact on the \emph{truncated} 
hyperspherical basis at a hypersphere of \emph{very large} radius. However on a 
 practical level, these two requirements contradict each other: the movement 
further in radius requires the increase of the basis size; a larger basis size 
may require a larger radius. Therefore, at some point, the further radial 
propagation of the solution (with fixed basis size) leads to a deterioration of 
its quality. For $^{45}$Fe with the decay energy of 1.154 MeV and the basis size 
of $K_{\max}=20$,  radii between 500 and 2000 fm are needed to get reasonable 
solutions.

There exists an analytical asymptotic of the three-body Coulomb problem (a so 
called ``Redmond-Merkuriev'' asymptotic \cite{ros73,mer77}), which is presumably 
applicable to the true three-body decay. Practical application of this 
asymptotic is technically complicated and it seems that there exists a very 
limited experience in  using such an asymptotic. At the moment we are going to 
avoid these complexities and to demonstrate that there exists a simple and 
practical way to treat the problem.


\section{Extrapolation along classical trajectories}
\label{sec:extrapol}


To perform a classical extrapolation of the quantum-mechanical result, we need 
to switch from a WF to classical trajectories. This should be made at some 
closed surface around the decay region. The procedure becomes especially simple 
if the whole surface is located in the region of classically allowed motion. 
Then the flux vectors at the surface can provide initial conditions for 
classical trajectories.

When using the HH coordinates there is only one variable $\rho$, which has a 
dimension of length [the 6-dimensional flux can be calculated for different 
$\rho$ values, see Eq.\ (\ref{eq:newton-flux})]. Therefore, it is natural in 
this approach to select the hypersphere with a large radius $\rho_{\max}$ as 
such a surface. We will see later that tiny regions on the hypersphere with a 
large radius, where the pairwise distances appear to be small, do not lead to 
problems as the WFs in these regions are strongly suppressed. This happens due 
to the energy conditions defining the true $2p$ decay: there are no long-living 
states in either pair of the three constituents and the strong Coulomb repulsion 
rapidly ``expels'' particles from the regions where they are close to each 
other.

A less evident, but important requirement is that the hyperradius $\rho_{\max}$ 
is large enough that the typical distances between each pair of particles 
significantly exceeds the typical quantum coherence length (the ``corpuscular'' 
aspect of the problem is then far prevailing over the possible wave effects). 
This is a complicated issue and in each case an acceptable minimal value of 
$\rho_{\max}$ should be defined by numerical experiment.

The classical trajectories formed at this hypersphere $\rho_{\max}$  are 
propagated to distances $\rho_{\text{ext}} \gg \rho_{\max}$ at which the 
momentum distributions are stabilized (what this exactly means we will see 
below). After this, the momentum distributions are reconstructed from the set of 
trajectories.

The pairwise distances, the Jacobi vectors, and the hyperradius are connected by
the following relations
\begin{eqnarray}
\mathbf{r}_{12}=\mathbf{X}\,, \quad \mathbf{r}_{23}=\mathbf{Y}-c_1 \mathbf{X}\,,
\quad \mathbf{r}_{31}=\mathbf{Y}+c_2 \mathbf{X} \,, \nonumber \\
\rho^2 =\frac{A_1 A_2}{A_1+A_2} X^2 + \frac{(A_1+ A_2)A_3}{A_1+A_2+A_3} Y^2 \,,
\nonumber \\
c_1 = A_1/(A_1+A_2) \,, \quad c_2 = A_2/(A_1+A_2) \,.
\label{eq:hh-coord}
\end{eqnarray}
In the definition of the hyperradius $\rho$, particle $A_3$ should be a heavy
core if $X$ and $Y$ are defined in the ``T'' Jacobi system and either $A_1$ or
$A_2$ should be a core in the ``Y'' Jacobi system  (see also Fig.\
\ref{fig:Jacobi} for the numbering convention).

The Newton equations of the motion for the Jacobi vectors are used to avoid
the extra degrees of the freedom connected to the center-of-mass motion:
\begin{eqnarray}
M_x \ddot{\mathbf{X}} & = & \frac{\alpha Z_1 Z_2 \mathbf{X}}{X^3}
- \frac{\alpha Z_2 Z_3 c_1 \mathbf{r}_{23}}{r_{23}^3}
+ \frac{\alpha Z_3 Z_1 c_2 \mathbf{r}_{31}}{r_{31}^3}   \,, \nonumber \\
M_y \ddot{\mathbf{Y}}  & = & \frac{\alpha Z_2 Z_3 \mathbf{r}_{23}}{r_{23}^3} +
\frac{\alpha Z_3 Z_1 \mathbf{r}_{31}}{r_{31}^3}  \,.
\label{eq:newton}
\end{eqnarray}
The particular choice of the form of Eqs.\ (\ref{eq:newton}) (``T'' or ``Y'' 
Jacobi
system) or the numerical precision in solving this system are not practical
obstacles for getting the correct classical trajectories.

The initial conditions for these equations are defined on the hypersphere of the
maximal radius achieved in the quantum-mechanical calculations:
\begin{eqnarray}
\{\rho_{\max},\Omega_{\rho}^{(r)} \}  \rightarrow  \{\mathbf{X}(0),\mathbf{Y}(0)
\} \,, \nonumber \\
\{ \mathbf{j}_x(\rho_{\max},\Omega_{\rho}^{(r)}),
\mathbf{j}_y(\rho_{\max},\Omega_{\rho}^{(r)}) \} \rightarrow
\{\dot{\mathbf{X}}(0),\dot{\mathbf{Y}}(0) \} \,,
\label{eq:newton-bc}
\end{eqnarray}
where $\Omega_{\rho}^{(r)}$ is a randomly generated 5-dimensional hyperangle
selected by the MC procedure according to the WF density $|\Psi_3^{(+)}|^2$ at
$\rho=\rho_{\max}$. The flux associated with the Jacobi vectors is defined in an
ordinary way:
\begin{equation}
\mathbf{j}_i(\rho, \Omega_{\rho})=\frac{1}{M_i}  \text{Im}
\left[\Psi_3^{(+)\dagger} \nabla_i  \Psi_3^{(+)}\right] \,.
\label{eq:newton-flux}
\end{equation}

In the quantum-mechanical model of the three-body decays
\cite{gri00b,gri01a,gri03c,gri07,gri07a}, the total flux $j$ through the
hypersphere $\rho=\rho_{\max}$ define the width
\begin{equation}
\Gamma = j/N\, ,
\label{eq:width}
\end{equation}
where $N$ is normalization of the WF $\Psi_3^{(+)}$ in the internal region. The 
momentum distribution (distribution density) is found as the differential of the 
flux $dj/[d\varepsilon \, d\cos (\theta_k)]$, see Eq.\ (\ref{eq:corel-param}). 
In this work we compare the quantum-mechanical distributions calculated at 
$\rho=\rho_{\max}$ (called below ``without classical extrapolation'' or 
``initial'') with distributions obtained by classical extrapolation to 
$\rho=\rho_{\text{ext}}$ (``with classical extrapolation'' or ``final'').


\subsection{Treatment of spins}


It is implied above that the flux is averaged over the initial spin states and
summed over the final spin states. Therefore the components of the WF
$\Psi_3^{(+)}$ with different total spin $S$ can be considered as different
``particles'' whose contributions to the total momentum distribution should be
added incoherently.

In general, three particles (or two Jacobi vectors) define a plane. Within this 
plane, the set of 6 equations (\ref{eq:newton}) can be reduced to 4 equations. 
However, the momentum vectors do not necessarily belong to this plane. It is 
evident that the geometry of the problem remains planar in the case of zero 
angular momenta of the $X$ and $Y$ subsystems (this situation is shown in Fig.\ 
\ref{fig:Jacobi}). For nonzero angular momenta, some additional considerations 
are required.

Let us consider the flux field induced by the ordinary two-body WF with $l \neq 
0$. For $m=0$, the flux is purely radial as the angular part of the WF $Y_{lm}$ 
is real (flux is an imaginary part of the gradient matrix element). For purely 
radial flux, the classical angular momentum associated with the particular 
trajectory is zero (radius and momentum vectors are collinear). This can be seen 
as a source of a confusion as the quantum-mechanical momentum of the WF and the 
classical momentum of the selected trajectory are explicitly different. The 
answer seems to be that the classical characteristic of the trajectory should be 
related to average corresponding characteristic of the WF.

In the three-body case, the ground-state WFs typically have two major 
components: the dominating $L=0$ component and an ``admixture'' $L=1$ component. 
We imply here that a spin-zero core is considered; the two spin 1/2 protons can 
then be coupled into the total spins $S=0$ or $S=1$. The $L=0$ component of the 
WF is formed by terms with angular momenta in the subsystem $l_x=l_y$. It is 
easy to check that the angular part of this WF $[Y_{l_x}\otimes Y_{l_x}]_{00}$ 
is real and thus the classical angular momentum associated with any trajectory 
induced by this WF is zero. The decay in this case is \emph{planar} (we mean 
that for any generated event, a plane can be selected in which the coordinate 
vectors and the momentum vectors of all three particles are simultaneously 
located).

It is more complicated when the $L=1$ component is considered. It is possible to 
demonstrate that for the $[Y_{l_x}\otimes Y_{l_x}]_{1M}$ component of the WF 
with $M=0$, the configuration of the classical momenta is planar, while for $M 
\neq 0$, the planes formed by the three radii and by the three momenta does not 
coincide. However according to the Wigner-Eckart theorem, in order to define the 
observables it is sufficient to calculate the matrix elements for only one 
projection and the rest are reconstructed by the angular momentum algebra. 
Therefore, it seems sufficient to calculate the distributions for $M=0$ (planar 
case calculations are especially simple), while the distributions for $M = \pm 
1$ should be the same.


\section{Test cases of solvable semianalytical models}


In Ref.~\cite{gri07}, a semianalytical model was developed which allows one to
treat exactly the asymptotic behaviour of the three-body Coulomb WF for certain
simplified three-body Hamiltonians. The basic idea of the model is that instead
of the real three-body Hamiltonian
\begin{equation}
H_{3}=T_x +T_y + V_{12}(\mathbf{r}_{12}) + V_{23}(\mathbf{r}_{23}) +
V_{31}(\mathbf{r}_{31}) \,,
\label{eq:h3}
\end{equation}
we use the model Hamiltonian depending not on pairwise vectors $\mathbf{r}_{ij}$
but on the Jacobi vectors $\mathbf{X}$ and $\mathbf{Y}$
\begin{eqnarray}
H_{3}=T_x + T_y +V_{x}(\mathbf{X}) + V_{y}(\mathbf{Y}) + V_{3}(\rho) \,,
\label{eq:h3m}
\end{eqnarray}
The three-body potential $V_{3}(\rho)$ is used in this work has the Woods-Saxon
form
\begin{eqnarray}
V_{3}(\rho)=V_{3}^{0}\left(1+\exp\left[ (\rho-\rho_{0})/a_{\rho} \right]
\right)^{-1}\;, \\
\rho_0 =  \sqrt{2} \; 1.2 \left(A_{\text{core}}+1\right)^{1/3} \;,
\label{eq:pot-3b}
\end{eqnarray}
with a small value of the diffuseness parameter $a_{\rho}=0.4$~fm. The depth
$V_{3}^{0}$ of this potential is used to control the decay energy of the
system. The potentials $V_{x}$ and $V_{y}$ contains the nuclear and the Coulomb
contributions. The Coulomb potential of the homogeneously-charged sphere with
a radius $r_{\text{sph}}$ is used. The nuclear parts are described by 
Woods-Saxon
formfactors with radii taken from systematics.

In conjunction with this simplified Hamiltonian of Eq.\ (\ref{eq:h3m}), we can
introduce an auxiliary Hamiltonian
\begin{equation}
\bar{H}_{3}=T_x +T_y + V_{x}(\mathbf{X}_{12}) + V_{y}(\mathbf{Y}_{23}) \,,
\label{eq:h3a}
\end{equation}
for which the Green's function can be constructed in analytical form
\begin{equation}
G_{E_{T}}^{(+)}(\mathbf{XY,X}^{\prime}\mathbf{Y}^{\prime})=\frac{1}{2\pi i}
\int_{-\infty}^{\infty} \!\!\! d E_x \,G_{E_x}^{(+)}(\mathbf{X,X}
^{\prime})\,G_{E_y}^{(+)}(\mathbf{Y,Y}^{\prime}) \,,
\label{eq:g3a}
\end{equation}
where $E_T$ is the total decay energy, $E_x=\varepsilon E_T$, and
$E_y=(1-\varepsilon) E_T$ are the energies of the Jacobi subsystems.
The above two-body Green's functions correspond to the $X$ and $Y$ 
subhamiltonians
of $\bar{H}_{3}$. Based on Eq.\ (\ref{eq:g3a}), the width and the energy
distribution for the system defined by the Hamiltonian of
Eq.\ (\ref{eq:h3m}) can be obtained from
\begin{equation}
\frac{d \Gamma}{d \varepsilon} =\frac{dj}{d\varepsilon}= \frac{8}{\pi} E_{T}
\frac{M_x M_y}{k_x(\varepsilon) k_y(\varepsilon)}\left|A(\varepsilon)\right|^2
\,,
\label{eq:dnde}
\end{equation}
where $dj/d \varepsilon$ is a differential of the flux at the asymptotic. For
a particular set of quantum numbers $l_x$, $l_y$, the amplitudes 
$A(\varepsilon)$ are
defined via the scattering eigenfunctions $\varphi_{l_i}$ of subhamiltonians of
(\ref{eq:h3a}):
\begin{eqnarray}
A(\varepsilon)=\int^\infty_0 dX \int^\infty_0 dY
\varphi_{l_x}(k_x(\varepsilon)X)
\, \varphi_{l_y}(k_y(\varepsilon)Y) \quad \nonumber \\
\times V_3(\rho)\, \varphi_{Ll_xl_yS}(X,Y) \,.\quad
\label{eq:ampl-sim}
\end{eqnarray}
The WF $\varphi_{Ll_xl_yS}(X,Y)$ is the quasistationary eigenfunction of
(\ref{eq:h3m}), deduced in a three-body hyperspherical approach. The particular
choice of the boundary conditions for this WF (for
sufficiently large radius of the ``box'') is not important in the model. The
quasistationary WF is normalized to unity in the internal region, which gives
the identity $d \Gamma/ d \varepsilon \equiv d j /d \varepsilon$ in Eq.\
(\ref{eq:dnde}).

The results obtained in this model are quoted below as ``exact'' as they do not
suffer from any convergence/stability issues. In Sections \ref{sec:direct} and
\ref{sec:dipro}, we will use  models with different simplified Hamiltonians to 
test
the classical-extrapolation procedure in the case of the $^{19}$Mg-ground-state
(g.s.) decay and only after that we will turn to more realistic situations.


\subsection{Direct-decay model}
\label{sec:direct}


The Hamiltonian of Eq.~(\ref{eq:h3m}) constructed in the ``Y'' Jacobi system 
corresponds to
some physically well justified approximations. Namely, (i) we neglect the
proton-proton interaction and (ii) for one of the core-proton potentials we use 
the
Jacobi $Y$ variable instead of the relative distance. The later assumption 
becomes correct
in the limit of an infinitely-heavy core and thus should
work well for heavy $2p$ emitters.

\begin{figure}[tb]
\centerline{
\includegraphics[width=0.240\textwidth]{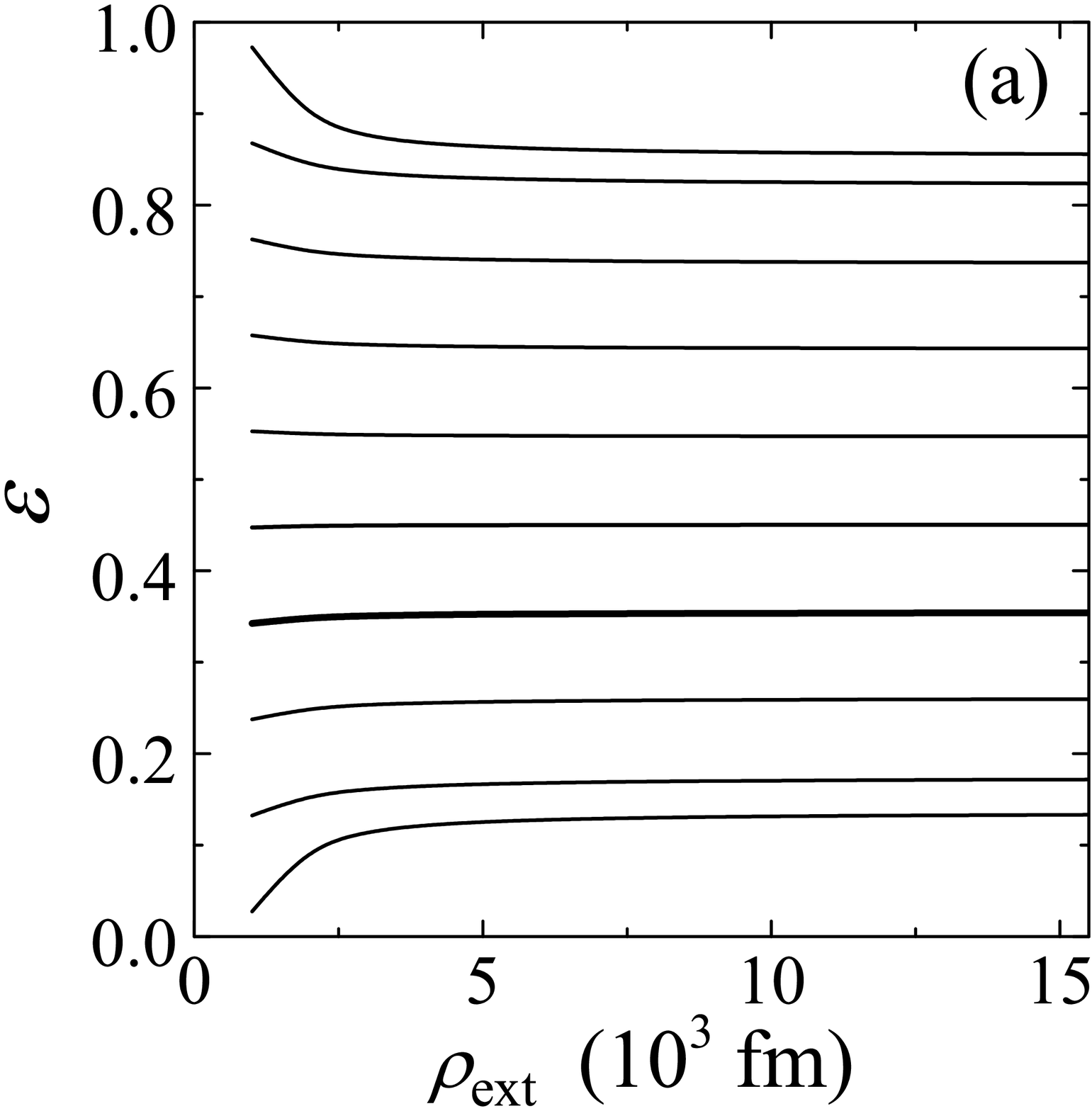}
\includegraphics[width=0.237\textwidth]{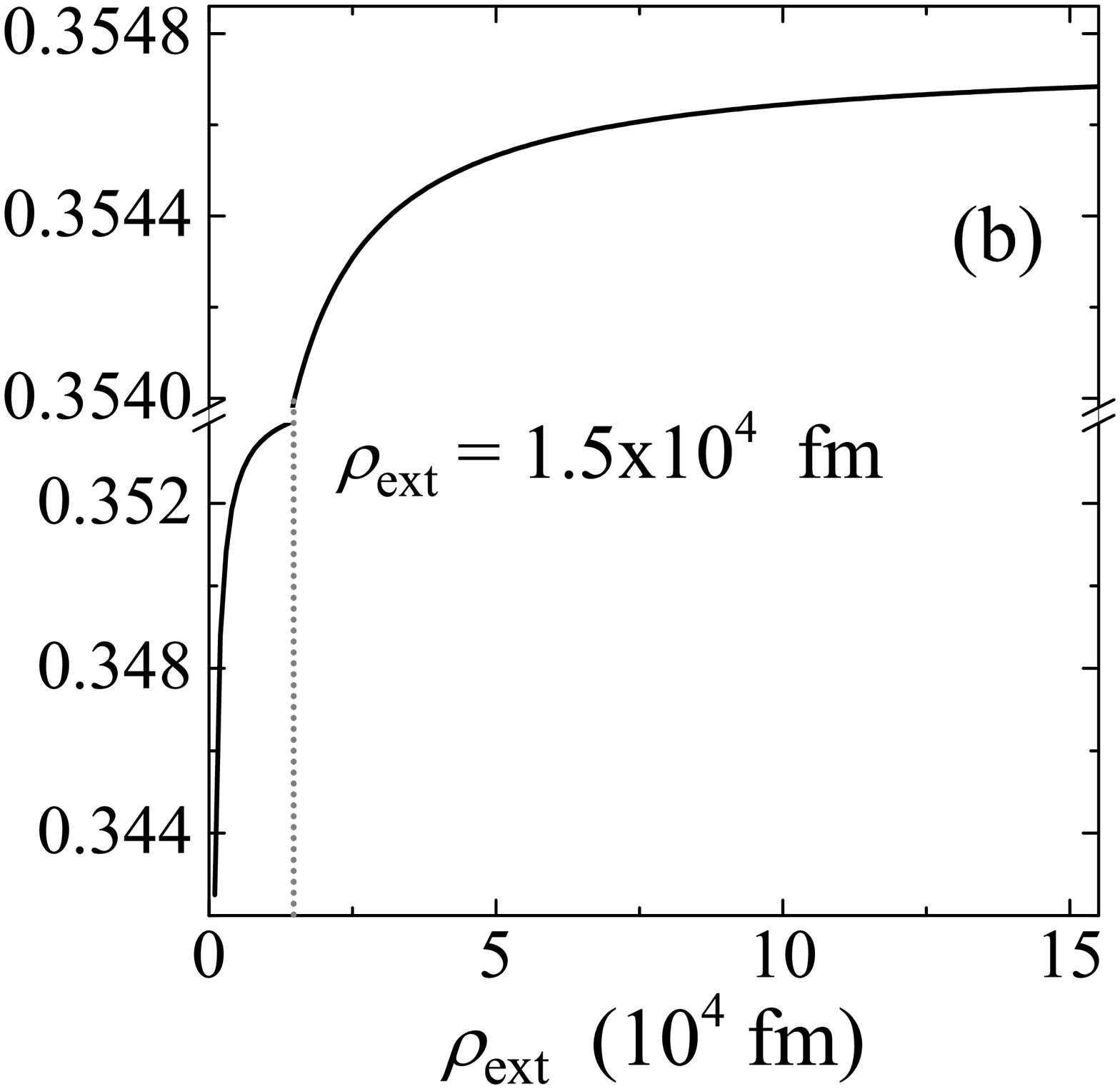}
}
\caption{ Classical trajectories for $^{19}$Mg in a direct decay
model ($\rho_{\max}=1000$ fm, $E_T=0.75$ MeV). Panel (b) shows one selected
trajectory on a large scale. The dotted line in panel (b) corresponds to the 
scale
of the panel (a).}
\label{fig:direct-traj}
\end{figure}

Let us consider the ``Y'' system, where the subsystem \{core+proton\} is taken
as an effective particle lying on the $X$ coordinate as shown Fig.\
\ref{fig:Jacobi}:
\begin{eqnarray}
V^{\text{coul}} = \frac{\alpha Z_1 Z_{2}}{X}+\frac{\alpha
(Z_1+Z_{2})Z_3}{Y}.
\label{eq:coul-3b-approx}
\end{eqnarray}
In this case, we include both pairwise interactions $V_x^{\text{nucl}}$ and
$V_y^{\text{nucl}}$. The system with such a composition of potentials in the 
``Y'' system was labeled as ``Two final-state interactions'' in Ref.\ 
\cite{gri07}.

For the $^{19}$Mg g.s., we assume the pure $d$-wave structure $l_x=l_y=2$  in 
this model. The nuclear Woods-Saxon potential was used with the radius
\begin{equation}
r_0 = 1.2 \left(A+1 \right)^{1/3}
\label{eq:sim-rad}
\end{equation}
and diffuseness $a=0.65$ fm. The depth of the potentials was adjusted to give
an energy of 1.3~MeV for the ground-state resonance in $^{18}$Na \cite{muk07} 
and the Coulomb potential of the charged sphere with radius
\begin{equation}
r_\text{sph} = \sqrt{\frac{5}{3}\left(1.2 A^{1/3}\right)^2+0.8^2}
\label{eq:sim-rad-sph}
\end{equation}
was used. In the above expressions, one should substituted $A=A_2$ in the $X$
subsystem and $A=A_2+1$ in the $Y$ subsystem. In this model, we obtained the
half-life of $T_{1/2}=58$ ps (corresponding to $\Gamma=7.9 \times 10^{-11}$ MeV) 
which is in qualitative agreement with the experimental value for $^{19}$Mg 
($T_{1/2}=4$ ps \cite{muk07}).

\begin{figure}[tb]
\includegraphics[width=0.4\textwidth]{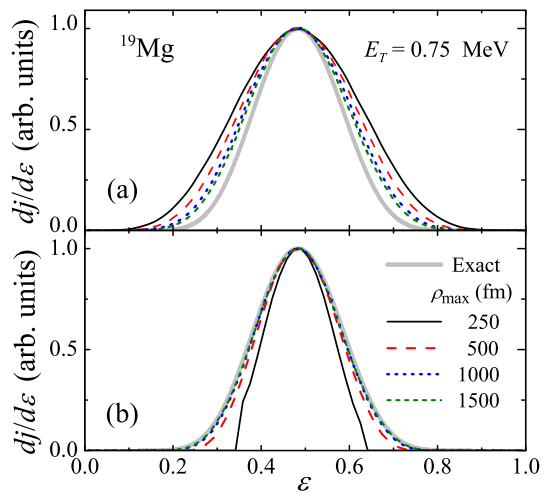}
\caption{(Color online) Energy distribution for $^{19}$Mg with different
$\rho_{\max}$ values without (upper panel) and with (lower panel) classical
extrapolation in the direct decay model. The calculations are performed with
$E_{\text{T}}=0.75$ MeV, $K_{\max}=20$, and lower panel with
$\rho_{\text{ext}}=40000$ fm. Gray curve shows the exact result of Eq.\
(\ref{eq:dnde}) (the same for
both panels).}
\label{fig:direct-radius}
\end{figure}

The radial convergence of the energy distribution $\varepsilon$ in this model 
for some classical trajectories is illustrated in Fig.\ \ref{fig:direct-traj}. 
The trend of the CE is to make the energy distribution narrower. The visual 
stability of the distributions is achieved at distances of about 
$\rho_{\text{ext}}\sim 7000$ fm [Fig.\ \ref{fig:direct-traj}(a)]. On a larger 
scale, a certain drift of the trajectories can be seen up to much larger 
distances [Fig.\ \ref{fig:direct-traj}(b)].

The effect of the CE on the energy distributions is demonstrated in Fig.\ 
\ref{fig:direct-radius}. The energy distributions have a characteristic bell 
shape. The upper panel shows the energy distributions calculated with the 
quantum-mechanical three-body model \cite{gri07} for different $\rho_{\max}$ 
values. The calculated result tends towards the ``exact'' result of Eq.\ 
(\ref{eq:dnde}), shown by the gray curves. However, this convergence is very 
slow and some discrepancy remains even for the largest available $\rho_{\max}$. 
The lower panel shows the distributions obtained with the classical 
extrapolation. These distributions are clearly wrong for $\rho_{\max} \lesssim 
500$ fm. However for larger $\rho_{\max}$, they stabilize and reproduce the 
results of the solvable model  [Eq.\ (\ref{eq:dnde})] within the width of the 
curve.


\subsection{``Diproton'' model}
\label{sec:dipro}

The word ``diproton'' in parenthesis is the name of this model as it is
different from the diproton model typically used in the literature. The diproton
correlation in our model is not introduced statically (which means ``by hand'')
but is treated dynamically. In Ref.~\cite{gri07}, we have demonstrated that
when introduced appropriately for the configurations with lowest possible
angular momenta in the subsystems, the diproton model can provide only a very
small value for the $2p$ width. For decays of the higher-$l$ configurations,
like $[p^2]$ or $[d^2]$ for $0^+$ states, this model overestimates the width.
Therefore, it is not applicable in practice, in contrast to widespread beliefs.

In this work we apply the diproton model, not for realistic estimates, but for
testing purposes. The diproton model  gives very sharp energy distributions
focused at small $p$-$p$ energies. So, we use it to determine whether the CE
procedure works for conditions of strong kinematical focusing.

\begin{figure}[tb]
\centerline{
\includegraphics[width=0.247\textwidth]{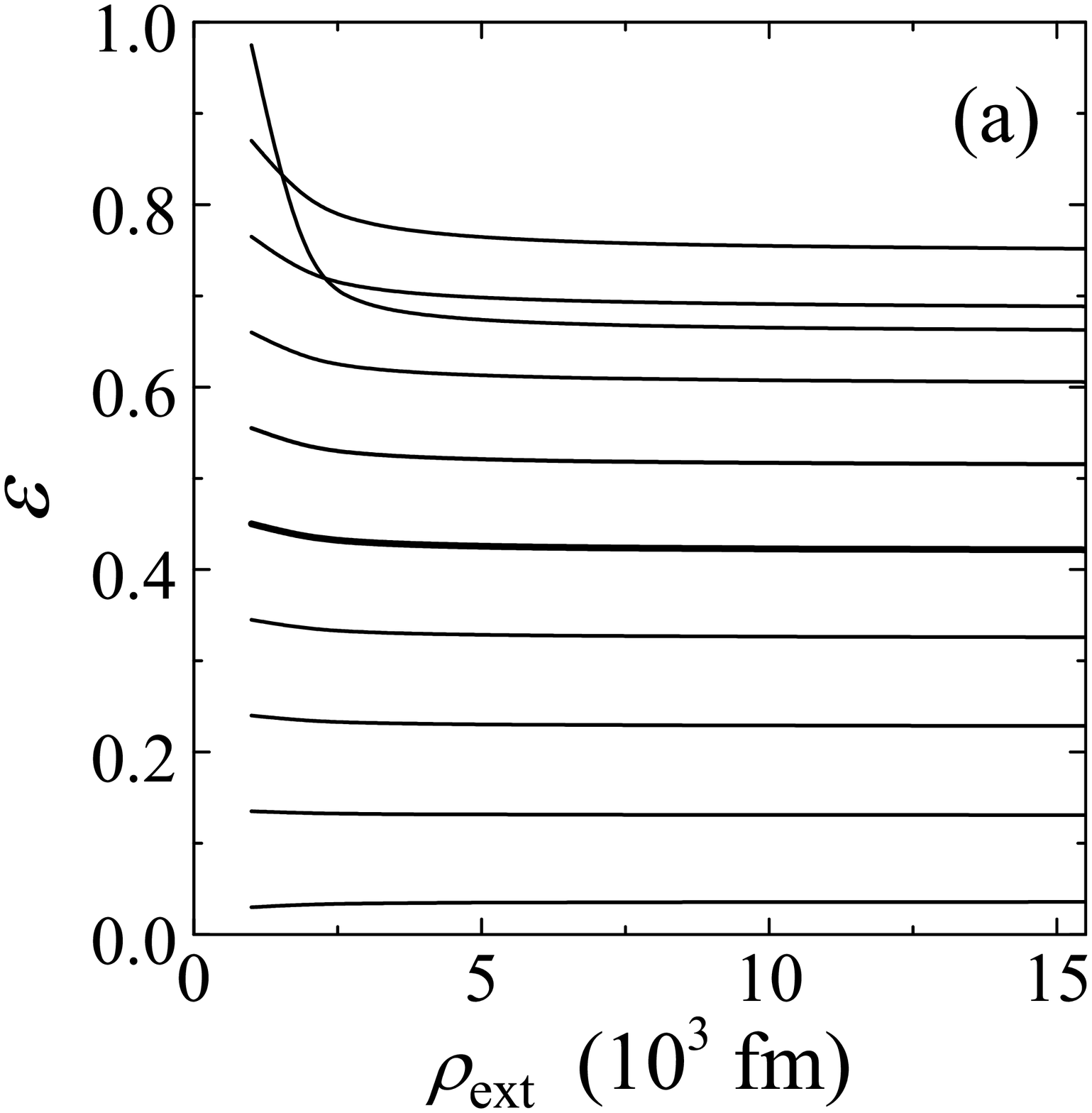}
\includegraphics[width=0.237\textwidth]{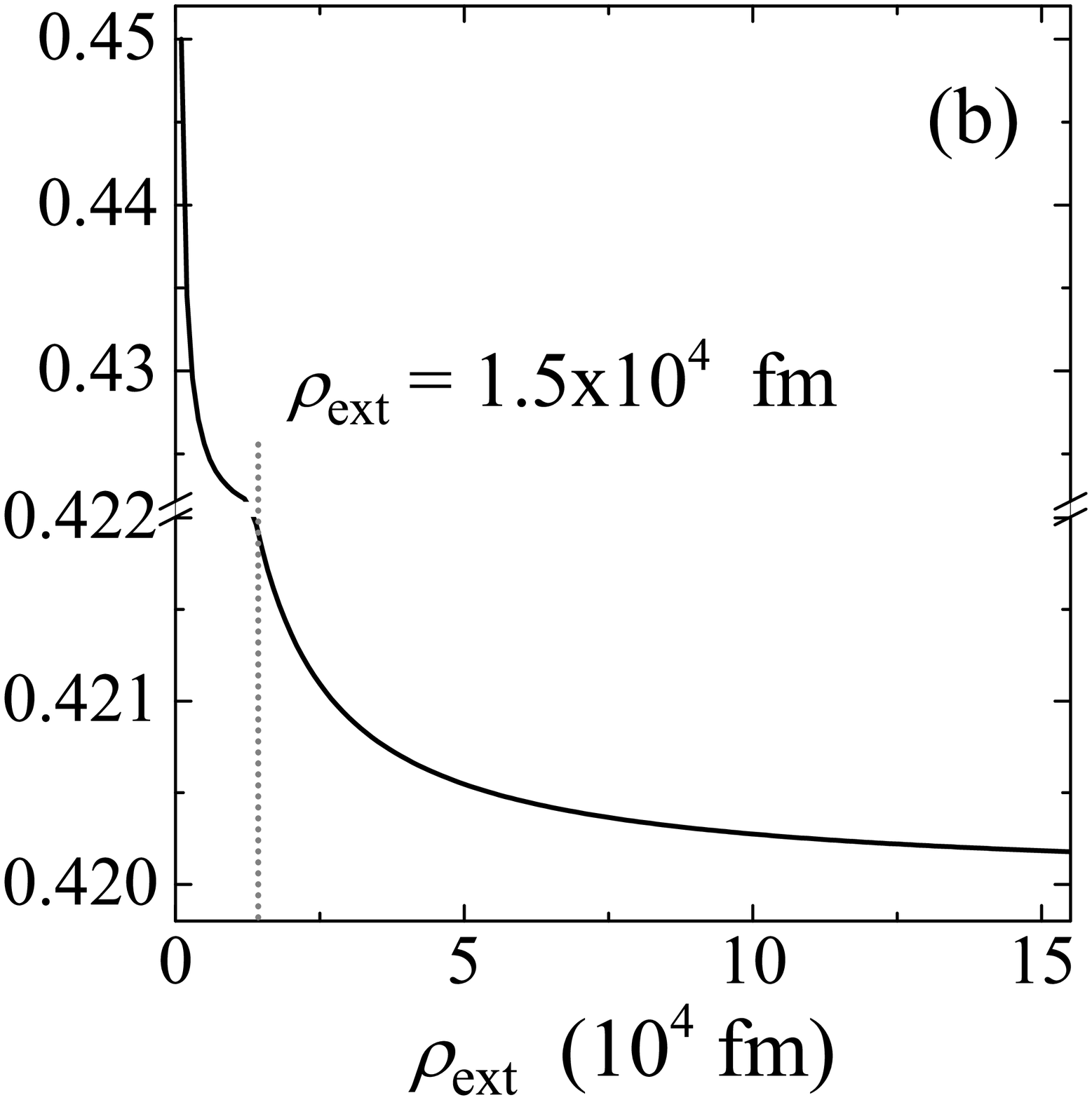}
}
\caption{ Classical trajectories for $^{19}$Mg in the ``diproton''
model ($\rho_{\max}=1000$ fm, $E_T=0.75$ MeV). Panel (b) shows one selected
trajectory on a large scale. The dotted line in panel (b) corresponds to the 
scale
of panel (a).}
\label{fig:dipro-traj}
\end{figure}

In the diproton model, Eqs.\ (\ref{eq:h3m})--(\ref{eq:ampl-sim})
are used in the ``T'' system, where the core $\{A_3,Z_3\}$ interacts
with the two protons as if they were an effective particle 
$\{A_1+A_2,Z_1+Z_2\}$. The
Coulomb potential of the simplified Hamiltonian can be written in the form
\begin{eqnarray}
V^{\text{coul}} =  \frac{\alpha Z_1 Z_2}{X}+\frac{\alpha
(Z_1+Z_2)Z_\text{core}}{Y}.
\label{eq:coul-pp}
\end{eqnarray}
Note, that this is a model with only one nuclear pairwise interaction
$V_x^{\text{nuc}}(X)$ in the $p$-$p$ channel (the second interaction can be put
to zero) and therefore the model is called ``One final-state interaction'' in
Ref.\ \cite{gri07}. The proton-proton nuclear potential for an $s$-wave
is taken as a single Gaussian
\begin{equation}
V(r) = V_0 \exp[-(r/r_0)^2] \,,
\label{eq:pot-pp}
\end{equation}
with $V_0=-31$ MeV and $r_0=1.8$ fm reproducing the low-energy $s=0$
nucleon-nucleon phase shifts. The Coulomb potential of the charged sphere with
radius
\begin{equation}
r_\text{sph} = \sqrt{\frac{5}{3} \left(1.2 A_{\text{core}}^{1/3}\right)^2 +
\frac{5}{3}
\left(1.2 \times 2^{1/3}\right)^2}
\label{eq:dipro-rad-sph}
\end{equation}
is used in the $Y$ coordinate. The half-life of $^{19}$Mg  obtained in this
model is $T_{1/2}=0.39$ ps (corresponding to $\Gamma=1.2 \times 10^{-9}$ MeV).

The radial convergence of the energy $\varepsilon$ in this model for some 
classical trajectories is illustrated in Fig.\ \ref{fig:dipro-traj}. The trend 
of the CE is for the trajectories to drift towards the more narrow ``diproton'' 
peak in the energy spectrum. The convergence trend is analogous to the 
direct-decay model with several thousand fm required for a reasonable 
stabilization and more than a hundred thousand fm required for complete 
stability.

\begin{figure}[tb]
\includegraphics[width=0.4\textwidth]{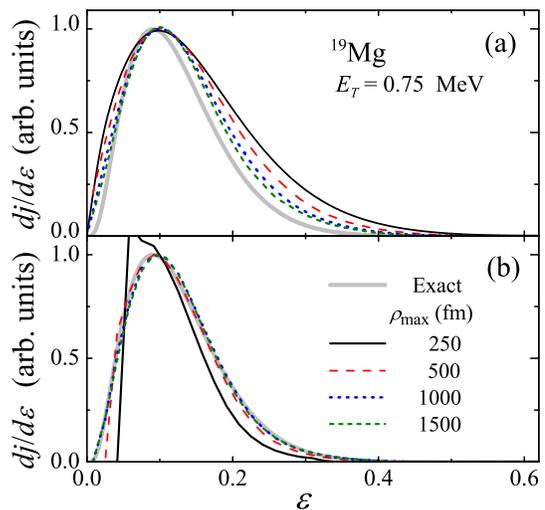}
\caption{(Color online) Energy distributions in $^{19}$Mg for the
``diproton'' model without (upper panel) and with (lower panel) classical 
extrapolation. Calculation results are shown for different $\rho_{\max}$ values. 
The calculations are performed with $E_{T}=0.75$ MeV, $K_{\max}=14$ and for 
lower panel with $\rho_{\text{ext}}=10^5$ fm. Gray curves show the ``exact'' 
result of Eq.\ (\ref{eq:dnde}) (the same for both panels).}
\label{fig:dipro-radius}
\end{figure}

The effect of the CE on the energy distribution is demonstrated
in Fig.~\ref{fig:dipro-radius}. The case appears to be completely
analogous to the direct-decay model. The upper panel shows the energy
distributions calculated within our three-body hyperspherical quantum-mechanical
approach for different  $\rho_{\max}$ values. The quantum-mechanical results
tend towards the ``exact'' result (\ref{eq:dnde}), but only very slowly. The
distributions provided by the classical extrapolation (see the lower panel in
Fig.\ \ref{fig:dipro-radius}) contain artifacts for $\rho_{\max}
\lesssim 500$ fm, but for larger $\rho_{\max}$, they stabilize and reproduce the
result of the solvable model Eq.\ (\ref{eq:dnde}) within the width of the curve.

\subsection{Brief conclusions}


Before we continue studies of realistic cases, let us outline what we can
conclude on the basis of the exactly-solvable models with simplified
Hamiltonians.

\begin{itemize}

\item[(i)] The quantum-mechanical calculations performed for $\rho_{\max}$ of
a few thousand fm give energy distributions which have visible deviations from
the ``exact'' results obtained in the semianalytical model. The extrapolated
distributions practically coincide with the ``exact'' ones.

\item[(ii)] The CE provides decent results only if the starting point for the
extrapolation is sufficiently large. Pragmatically, this means that the 
classical
trajectories in the kinematical space $\{\varepsilon,\cos(\theta_k)\}$ should be
quite short. The same should be true in the conjugated coordinate space. It can
be expected that the criterion of a successful transition from quantum to
classical calculation is that the classical variation of a position in some
space should be smaller than corresponding coherence length.

\item[(iii)] Distances of tens of thousands of fm's are needed to achieve 
complete stabilization of classical trajectories in practice. Some very minor 
drift continues after that, reflecting the long-range nature of the Coulomb 
interaction. However, it is evident that distances of $\sim 10^5$ fm are already 
atomic scale distances and the nuclear Coulomb effects should be suppressed for 
larger distances due to some form of electron screening.

\end{itemize}

Near perfect convergence of the extrapolated distributions to those calculated 
in the exact semianalytical models with the simplified Hamiltonians is \emph{not 
a proof} that the procedure should work perfectly in the case of a complete 
three-body Hamiltonian. However, it is very encouraging and we can expect that 
the quality of convergence in the realistic case will be very similar, as the 
kinematical conditions for the decay in the simplified models are chosen to be 
the same as in the realistic cases.


\section{Realistic three-body cases}


For the models with simplified Hamiltonians, we demonstrated only the energy 
distributions (angular distributions are trivial) and only in one Jacobi system 
(the one in which the particular semianalytical model is formulated). Conversion 
of the distribution into the other Jacobi system in this case does not provide 
an additional information. For realistic calculations we will demonstrate 
complete correlation pictures (on kinematical $\{\varepsilon, \cos(\theta_k) \}$ 
plane) simultaneously in both ``T'' and ``Y'' Jacobi systems. It should be 
understood that correlation pictures in ``T'' and ``Y'' Jacobi systems are just 
different representations of the same physical phenomenon. Conversion between 
these distributions is trivial. Nevertheless, we systematically demonstrate both 
of them simultaneously as each representation allows one to reveal different 
aspects of the correlations (see Ref.\ \cite{gri09} as example).


\subsection{Decay of the $^{6}$Be}


Very precise complete correlation data have recently been obtained
for $^{6}$Be in Ref.\ \cite{gri09d}.
The detailed theoretical studies of $2p$ decay of
the $^{6}$Be $0^+$ ground state have been carried out in this work and compared
to the experimental data. The dynamical range of around $\rho_{\max}=1000$ fm
used in these calculations was estimated in Ref.\ \cite{gri09d} as sufficient
for essentially complete convergence of the momentum distributions. A very nice
agreement between theory and experiment was found in this work. We would like to
check here whether the conclusions obtained in Ref.\ \cite{gri09d} can be
influenced by a more careful treatment of the momentum distributions.

\begin{figure}[tb]
\begin{tabular}{cc}
\includegraphics[width=0.251\textwidth]{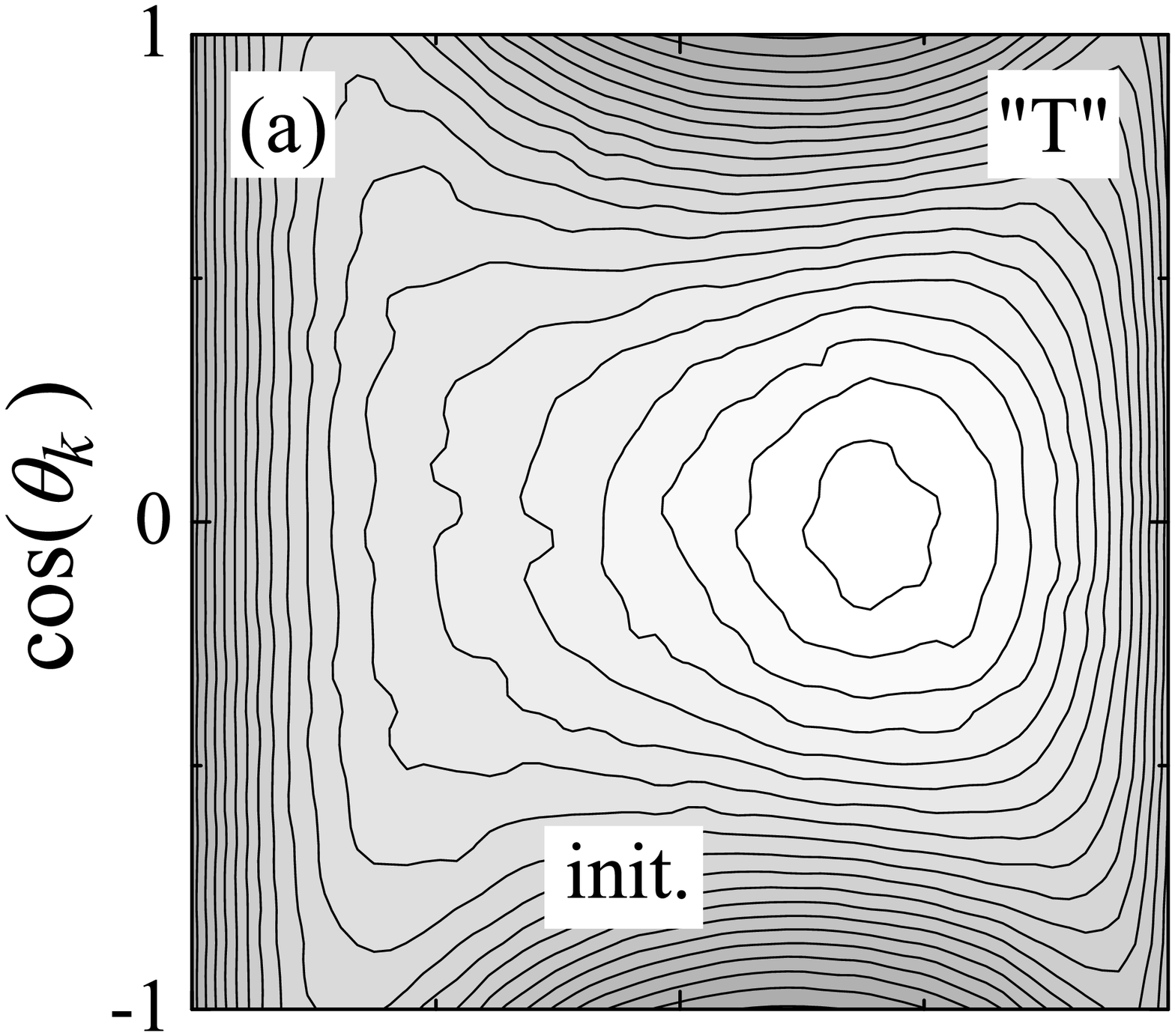}   &
\includegraphics[width=0.217\textwidth]{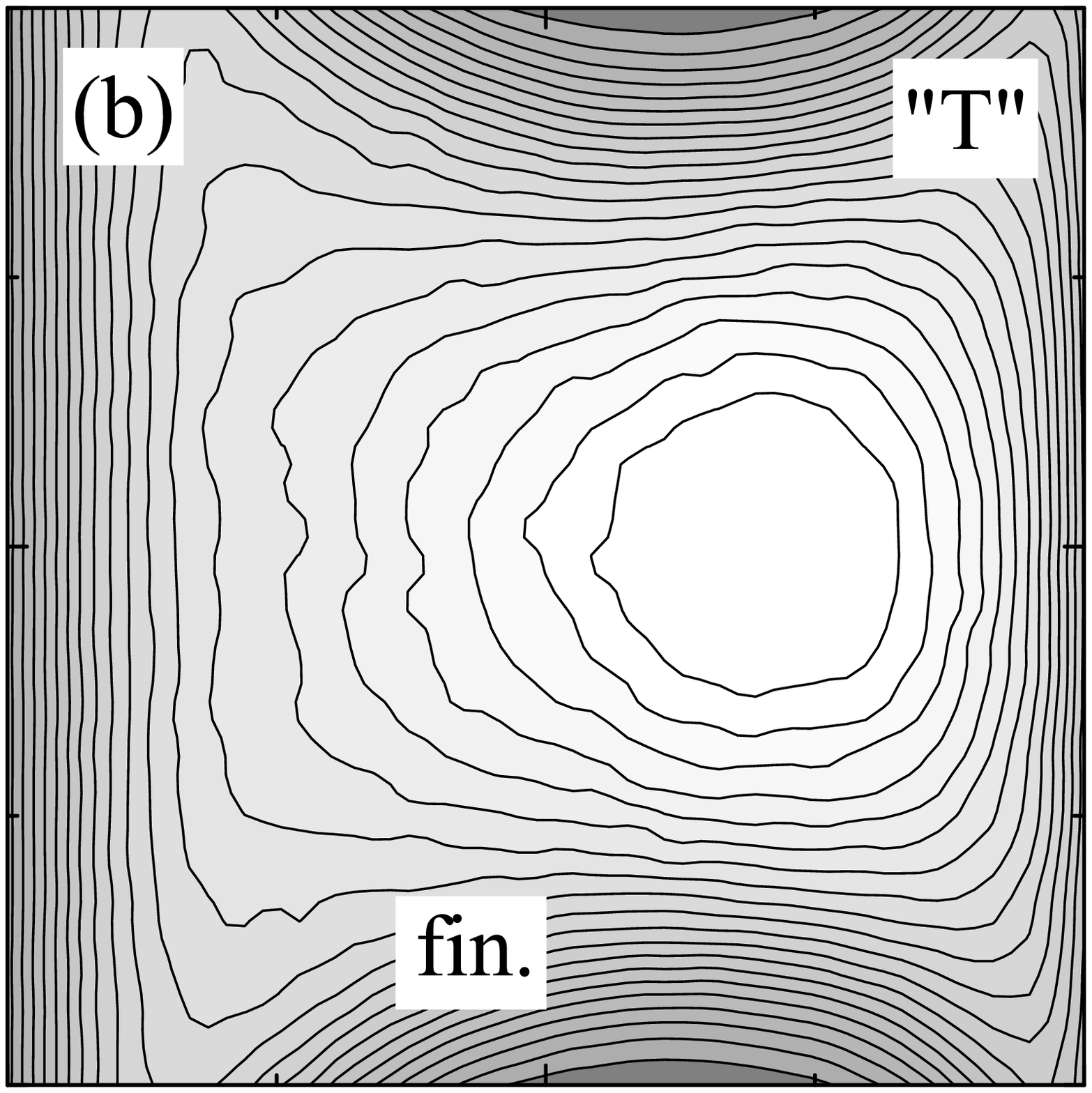} \\
\includegraphics[width=0.251\textwidth]{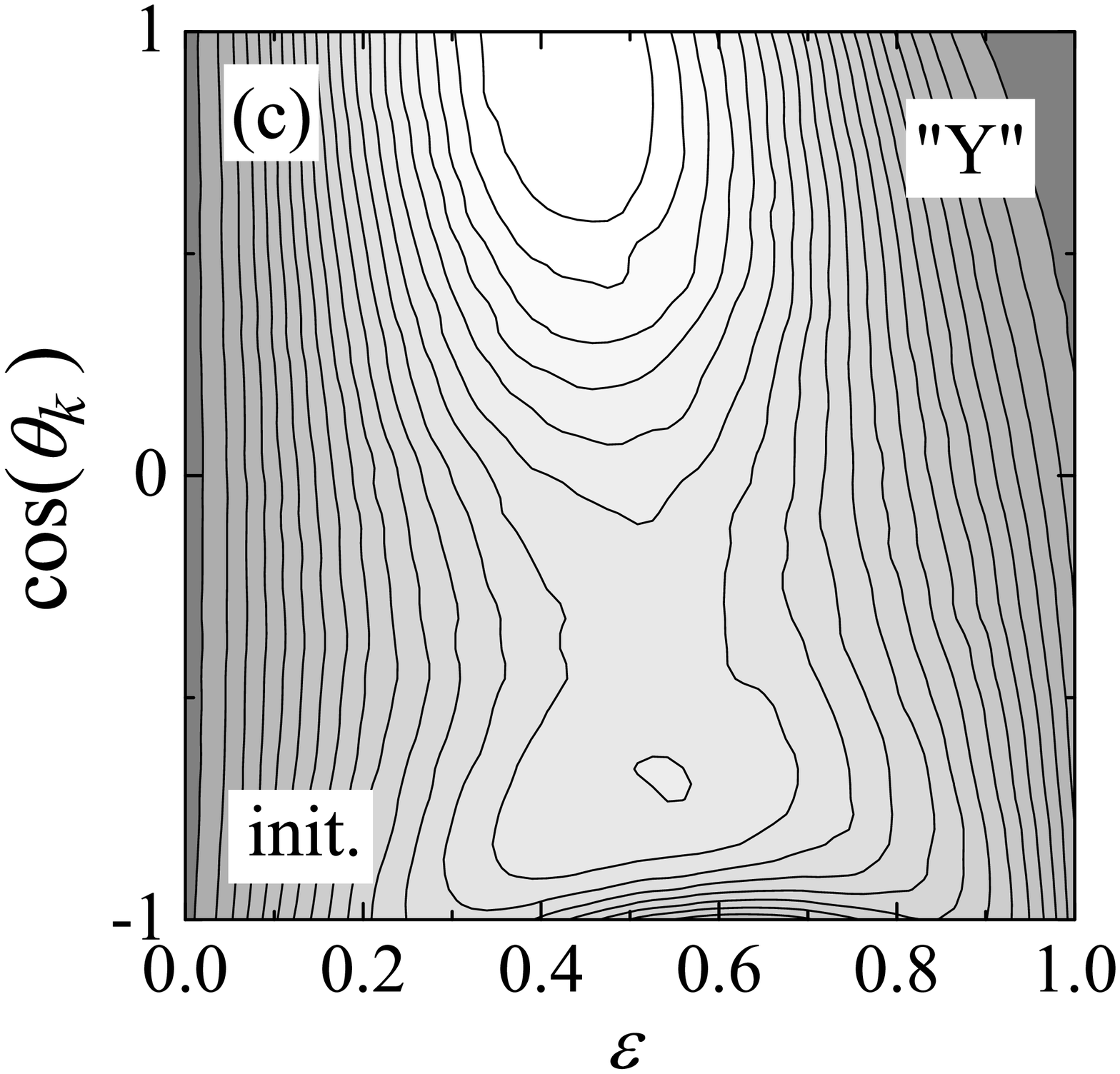}    &
\includegraphics[width=0.217\textwidth]{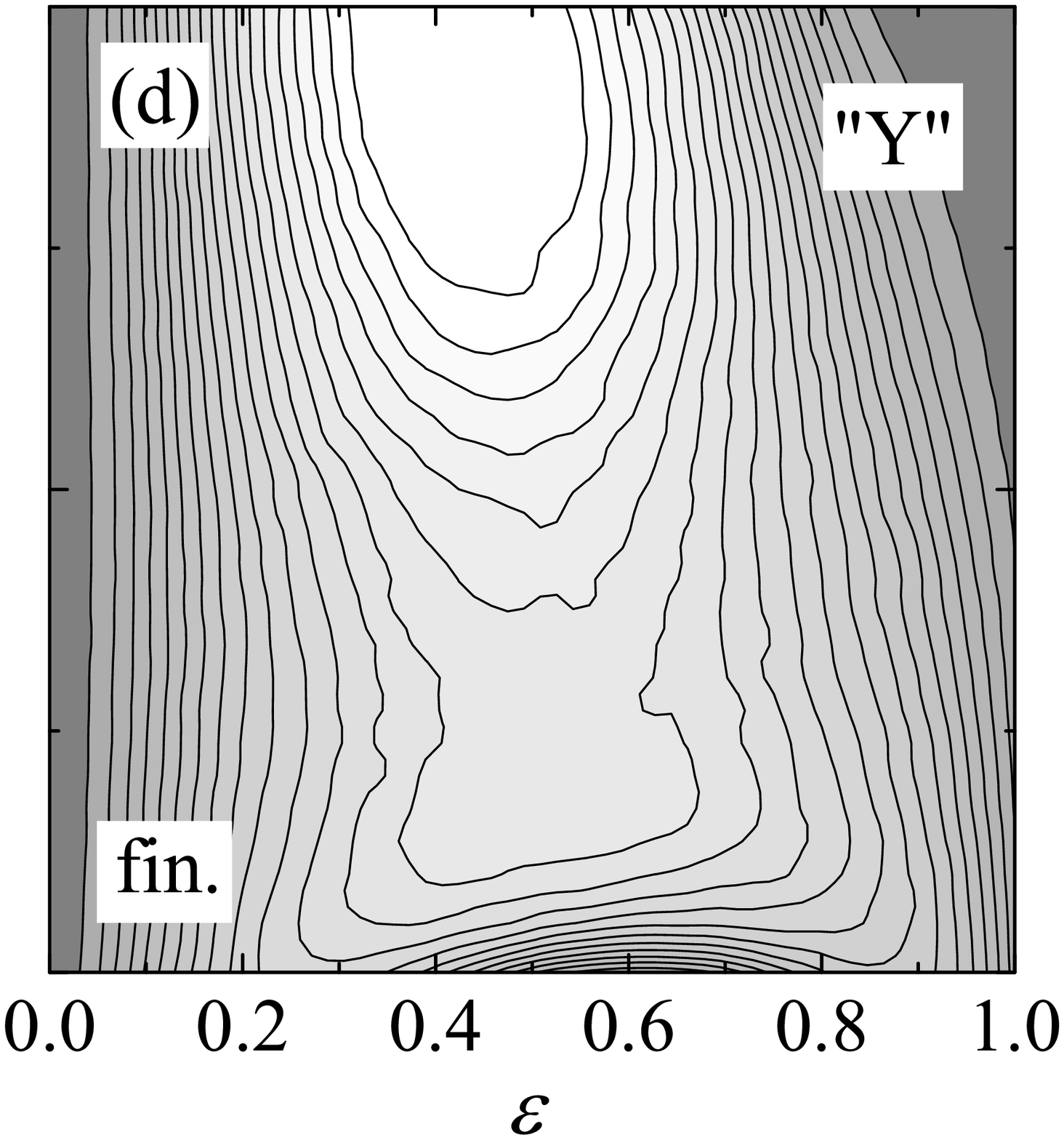}
\end{tabular}
\caption{Contour maps of the distribution density on the kinematical plane
$\{\varepsilon,\cos (\theta_k ) \}$ for $^{6}$Be in ``T'' (upper row) and ``Y''
(lower row) Jacobi coordinate systems without (left panels, ``init.'') and with
(right panels, ``fin.'') classical extrapolation.}
\label{fig:6be-complete}
\end{figure}

\begin{figure}[tb]
\begin{tabular}{ll}
\includegraphics[width=0.24\textwidth]{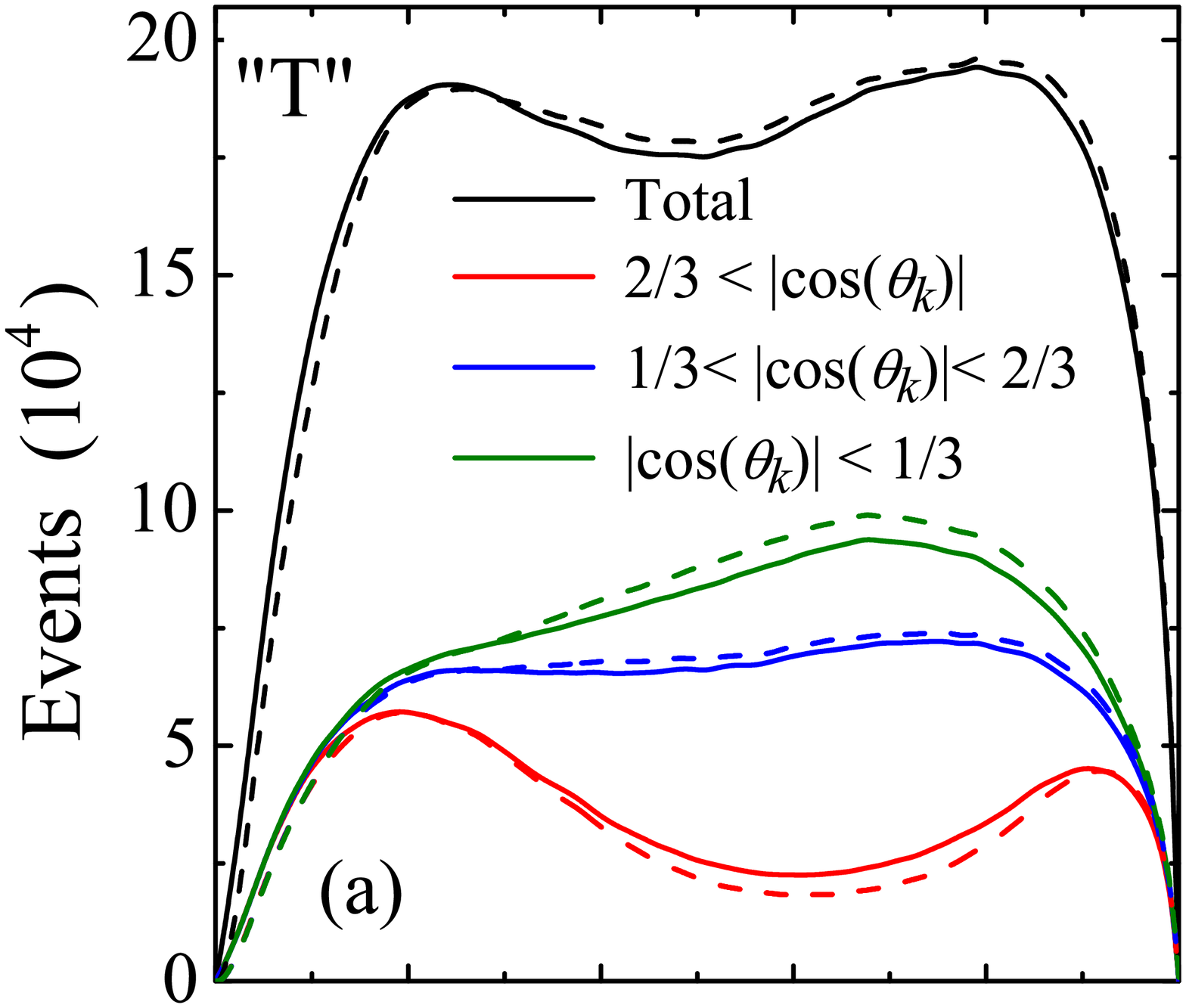}   &
\includegraphics[width=0.214\textwidth]{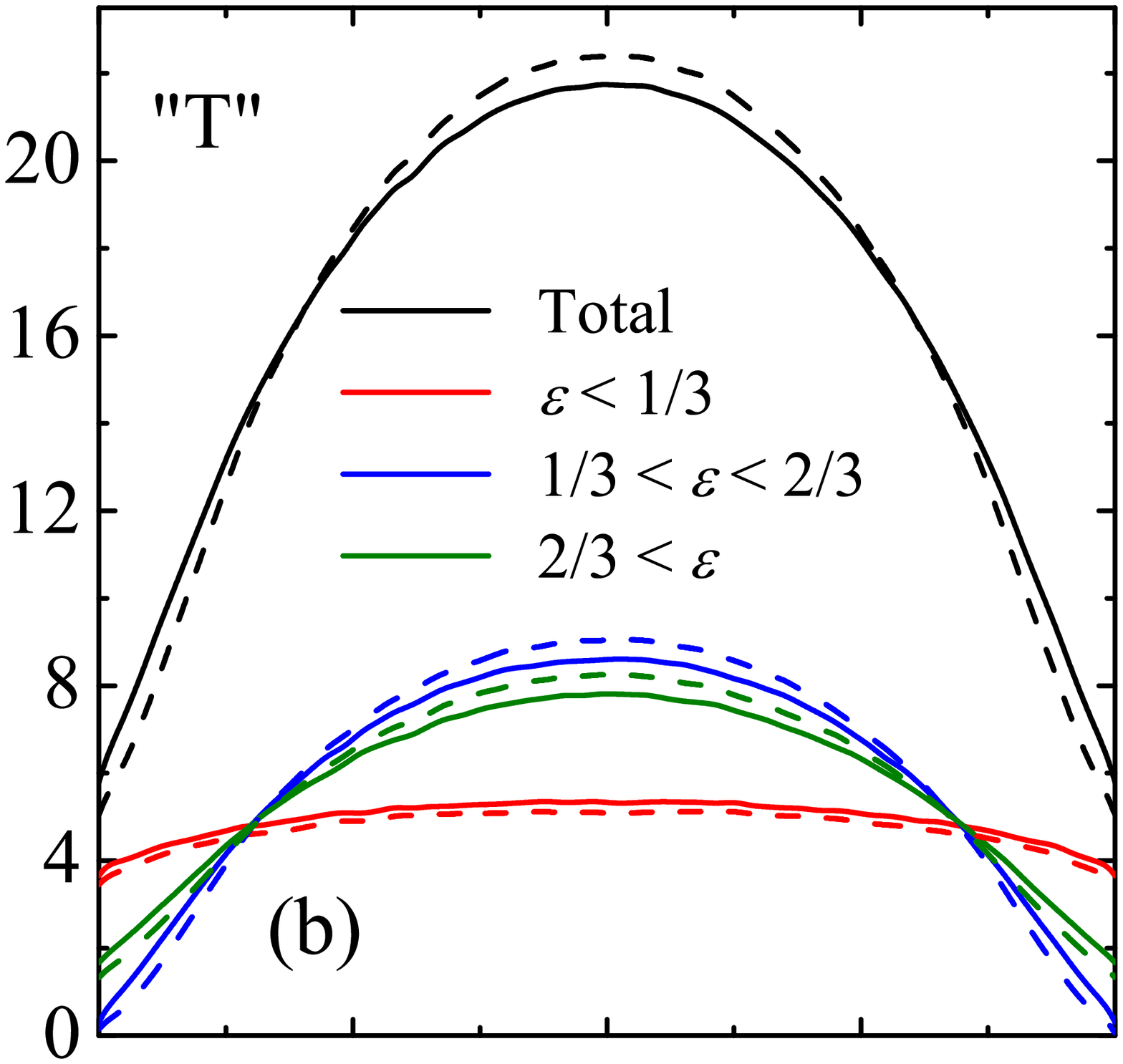} \\
\includegraphics[width=0.25\textwidth]{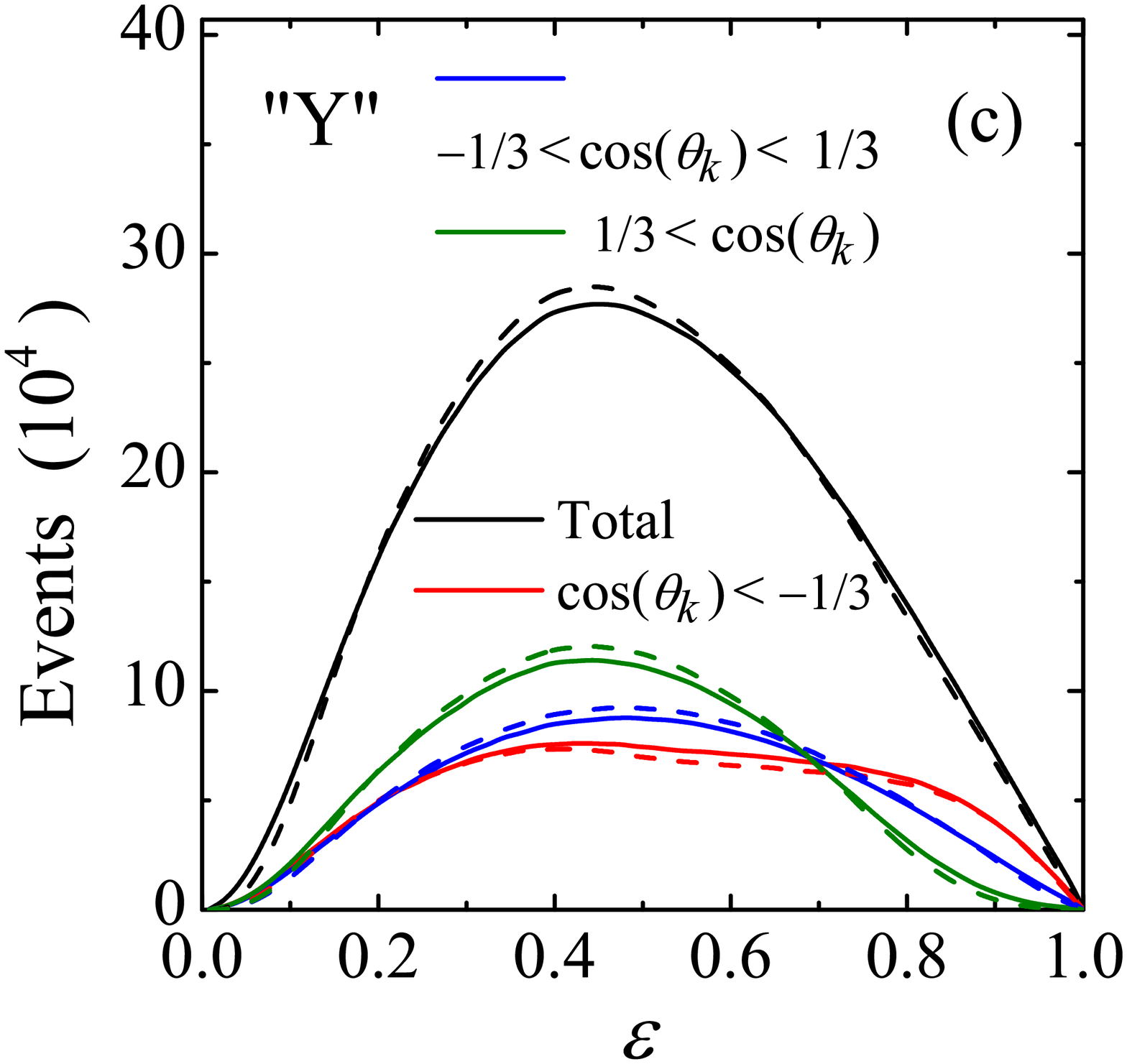}    &
\includegraphics[width=0.226\textwidth]{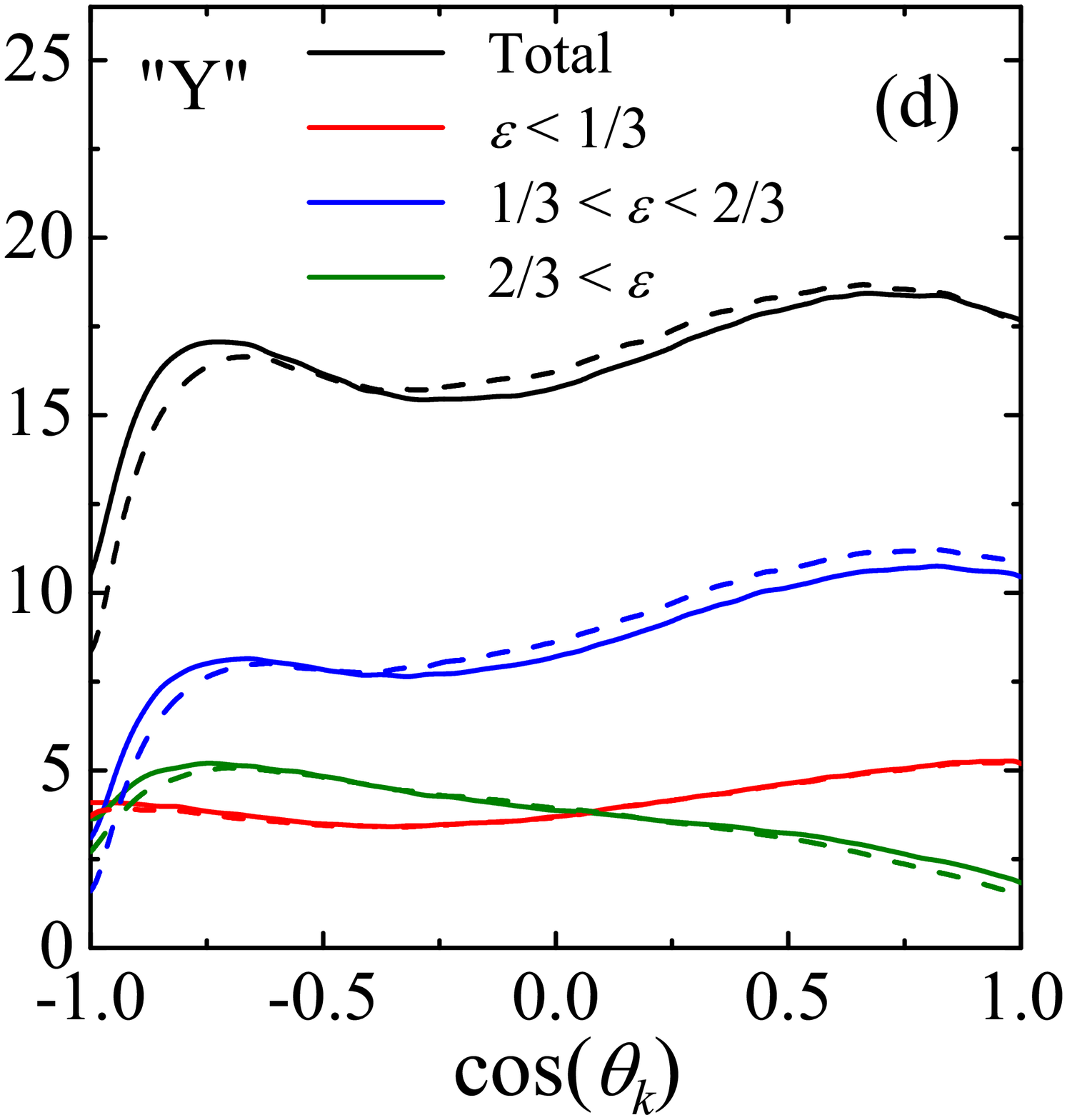}
\end{tabular}
\caption{(Color online) Inclusive energy and angular distributions for $^{6}$Be
in ``T'' (upper row) and ``Y'' (lower row) Jacobi coordinate systems without
(solid curves) and with
(dashed curves) classical extrapolation. Black lines
show the total distribution and the color coded lines show the inclusive
distributions for certain energy and angular bins (described in the legends).}
\label{fig:6be-proj}
\end{figure}

The classical trajectories for $^{6}$Be in the kinematical space are all very
short. Only the trajectories corresponding to small initial inter-particle
distances [$\varepsilon \sim 0.5$, $\cos(\theta_k) \sim \pm 1$ in ``T'' system]
have noticeable lengths. The complete correlation densities without and with
extrapolation are shown in Fig.\ \ref{fig:6be-complete} (this is the calculation
with potential set P2 from Ref.\ \cite{gri09d}, which was found to be the
optimal choice in that work). The distributions are very similar except for the
aforementioned regions of small initial inter-particle distances. A closer look 
at
these regions is provided in the inclusive distributions in Fig.\
\ref{fig:6be-proj}. The maximal effect can be found at small $\varepsilon$
values (corresponding to the lowest relative-energy motion between two of the
particles) or for the angular distribution in the middle energy bins around
$\cos(\theta_k)\sim \pm 1$ (in the ``T'' system) and $\cos(\theta_k)\sim - 1$
(in the ``Y'' system).

\begin{figure}
\includegraphics[width=0.43\textwidth]{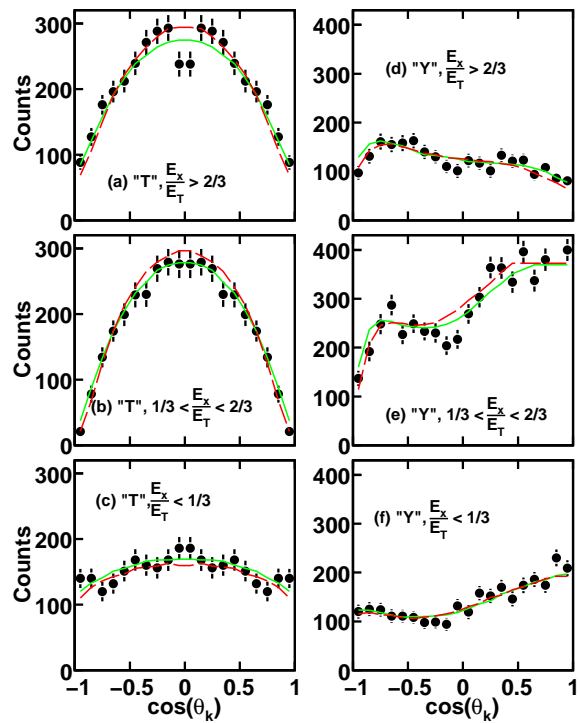}
\caption{(Color online) Comparison of experimental (data points \cite{gri09d})
and predicted (curves) $\cos(\theta_k)$ distributions in the ``T'' (left) and
``Y'' (right) Jacobi systems for the indicated gates on $\varepsilon= E_x/E_T $
parameter. The solid (green) and dashed (red) curves correspond to the
three-body calculations without and with classical extrapolation. The effect of
the detector bias and resolution is included.}
\label{fig:6be-exp}
\end{figure}

Comparisons with experimental angular distributions \cite{gri09d} are shown in
Fig.\ \ref{fig:6be-exp}. The theoretical curves here are visibly distorted
(relative to Fig.\ \ref{fig:6be-proj}) as the comparison is based on the full MC
simulation of the experimental setup \cite{gri09d}, which takes into account the
effects of the experimental bias and resolution. The effect of the classical
extrapolation is at the limit of the experimental sensitivity. Quantitatively 
the
$\chi^2/\nu$ values without extrapolation are 1.17 (in the ``T'' system) and
1.14 (in the ``Y'' system). The same values with extrapolation are found as 1.20
and 1.16, respectively. This is a little worse, but not really significant.
On the other hand, there seems to be a minor improvement of the agreement for 
the
parts of the middle energy bins mentioned in the previous paragraph.

The properties of the $^{6}$Be continuum are actively investigated now. The new
higher precision experiments have been performed recently at NSCL (Michigan
State University) and at Flerov laboratory (JINR, Dubna, Russia). The expected
precision of these experiments would make the improvement of the theoretical
distributions introduced in this work a necessary part of the data
interpretation.

\subsection{Decay of the $^{19}$Mg}

\begin{figure}[tb]
\includegraphics[width=0.42\textwidth]{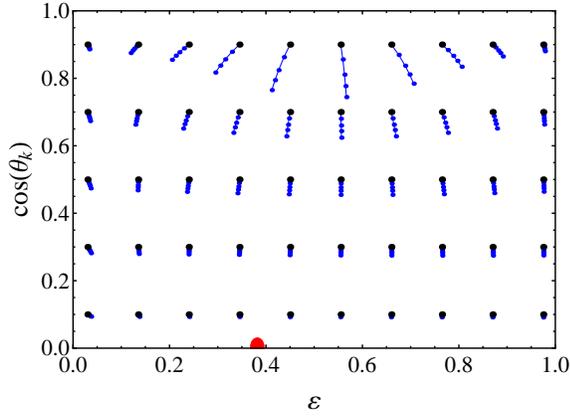}
\caption{(Color online)  Classical trajectories on the kinematical plane
$\{\varepsilon, \cos (\theta_k)\}$ for $^{19}$Mg in the Jacobi ``T'' system.
Starting points (larger black dots) correspond to $\rho_{\max}=1000$ fm. Dots in
the curves correspond to $\rho_{\text{ext}}$ equal 1300, 2000, 3500, and $10^5$
fm.  The red dot at the axis $\cos(\theta_k)=0$ corresponds to a stationary
point, see the discussion in Sec.\ \ref{sec:ss-solutions}.}
\label{fig:mg-traj}
\end{figure}

A systematic view of the classical trajectories on the kinematical plane for
$^{19}$Mg is given in Fig.\ \ref{fig:mg-traj}. The ``lengths'' of the 
trajectories
here are significant:  typically around $10-15 \%$ of the kinematical
variable range thus making the CE procedure necessary for quantitative
calculations of the momentum distributions.

\begin{figure}[tb]
\begin{tabular}{cc}
\includegraphics[width=0.251\textwidth]{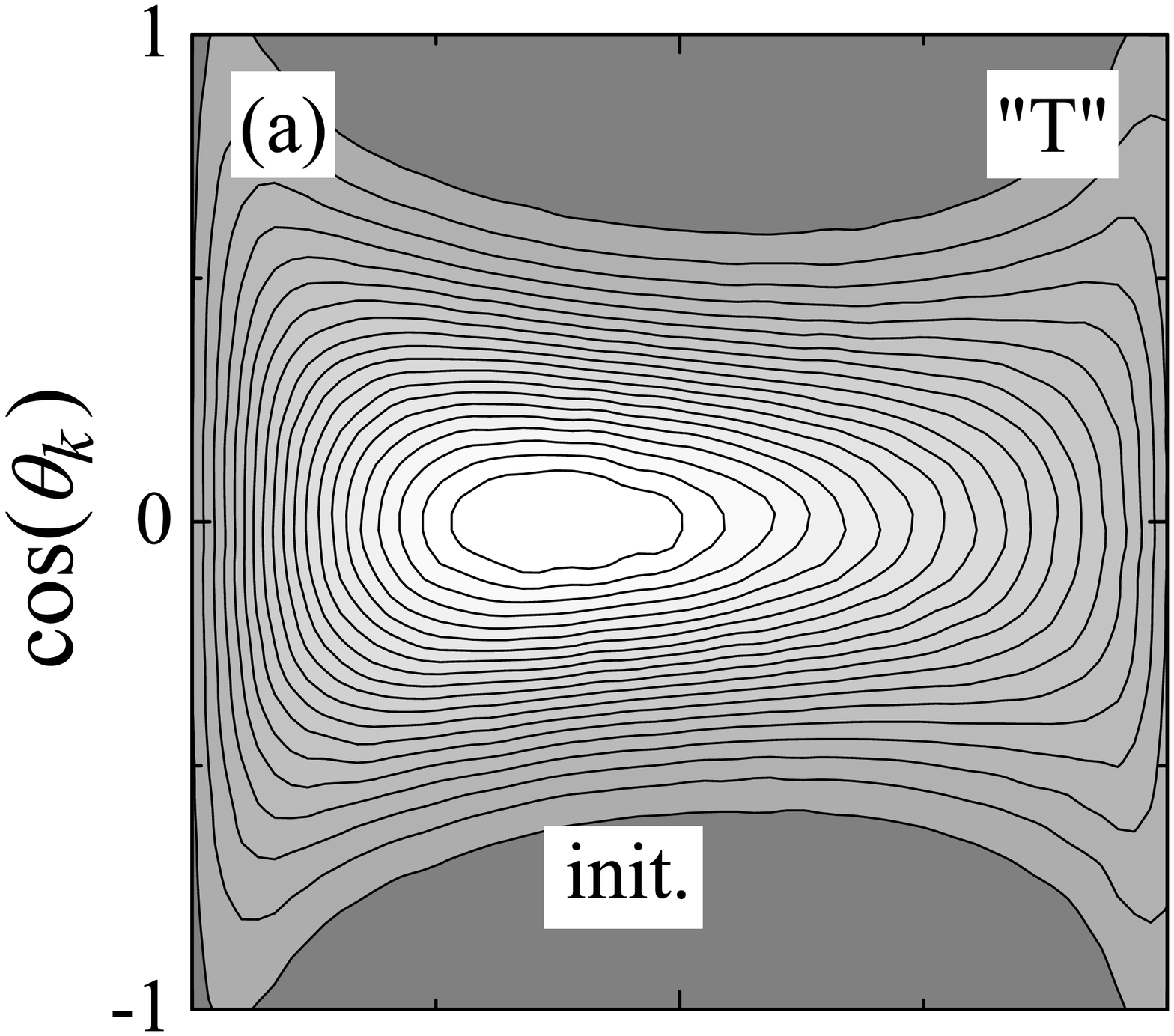}   &
\includegraphics[width=0.217\textwidth]{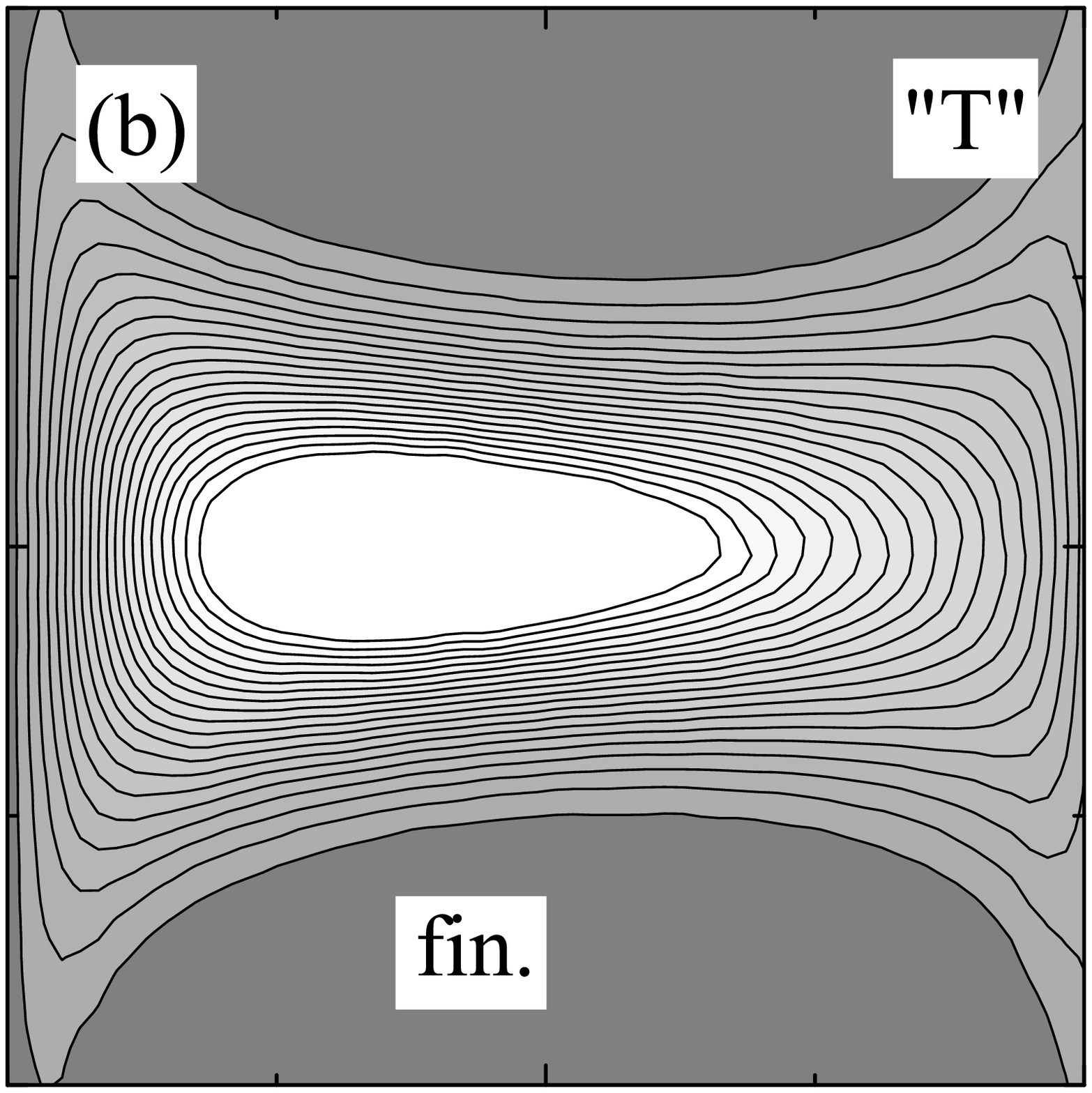} \\
\includegraphics[width=0.251\textwidth]{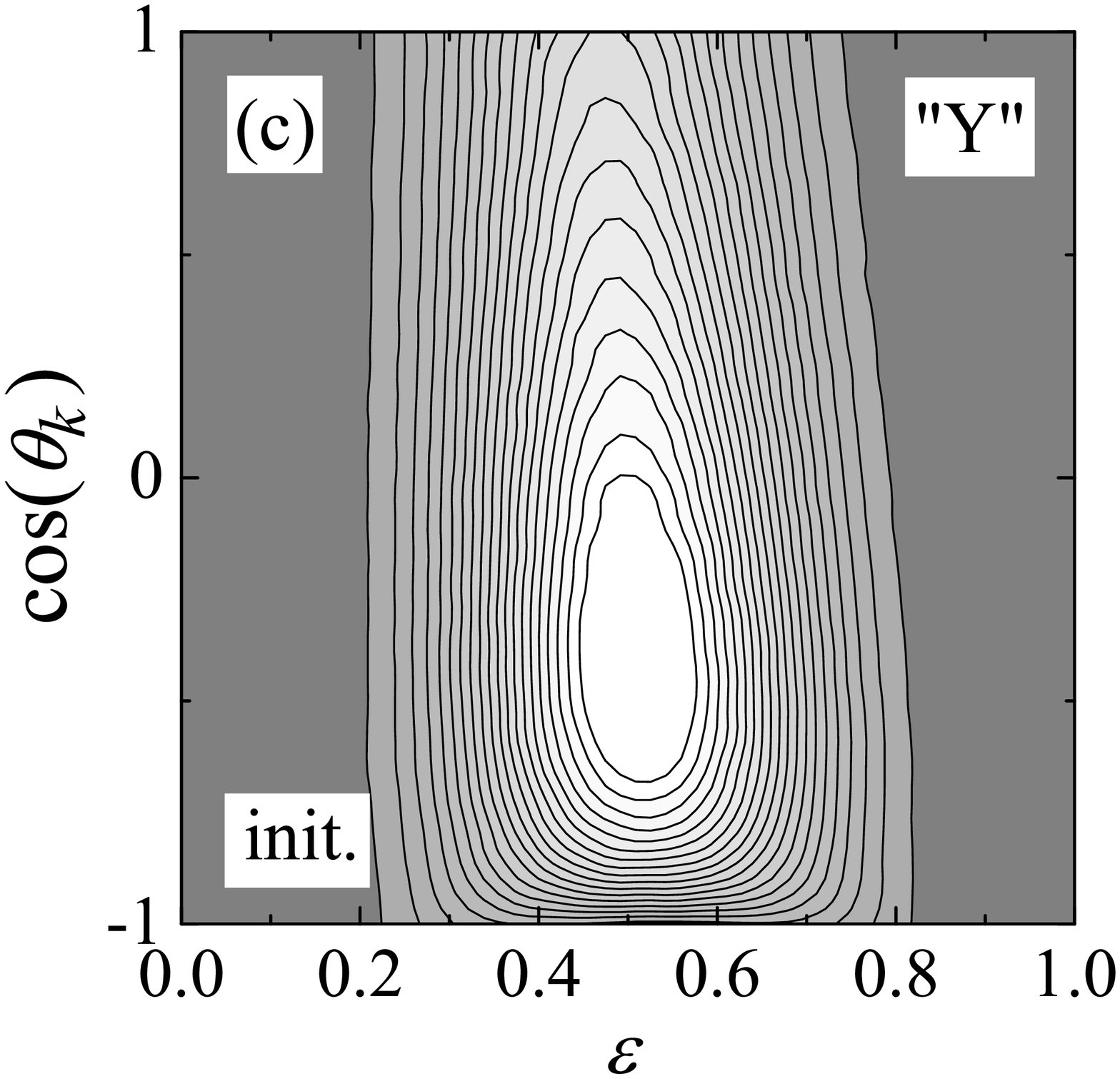}    &
\includegraphics[width=0.217\textwidth]{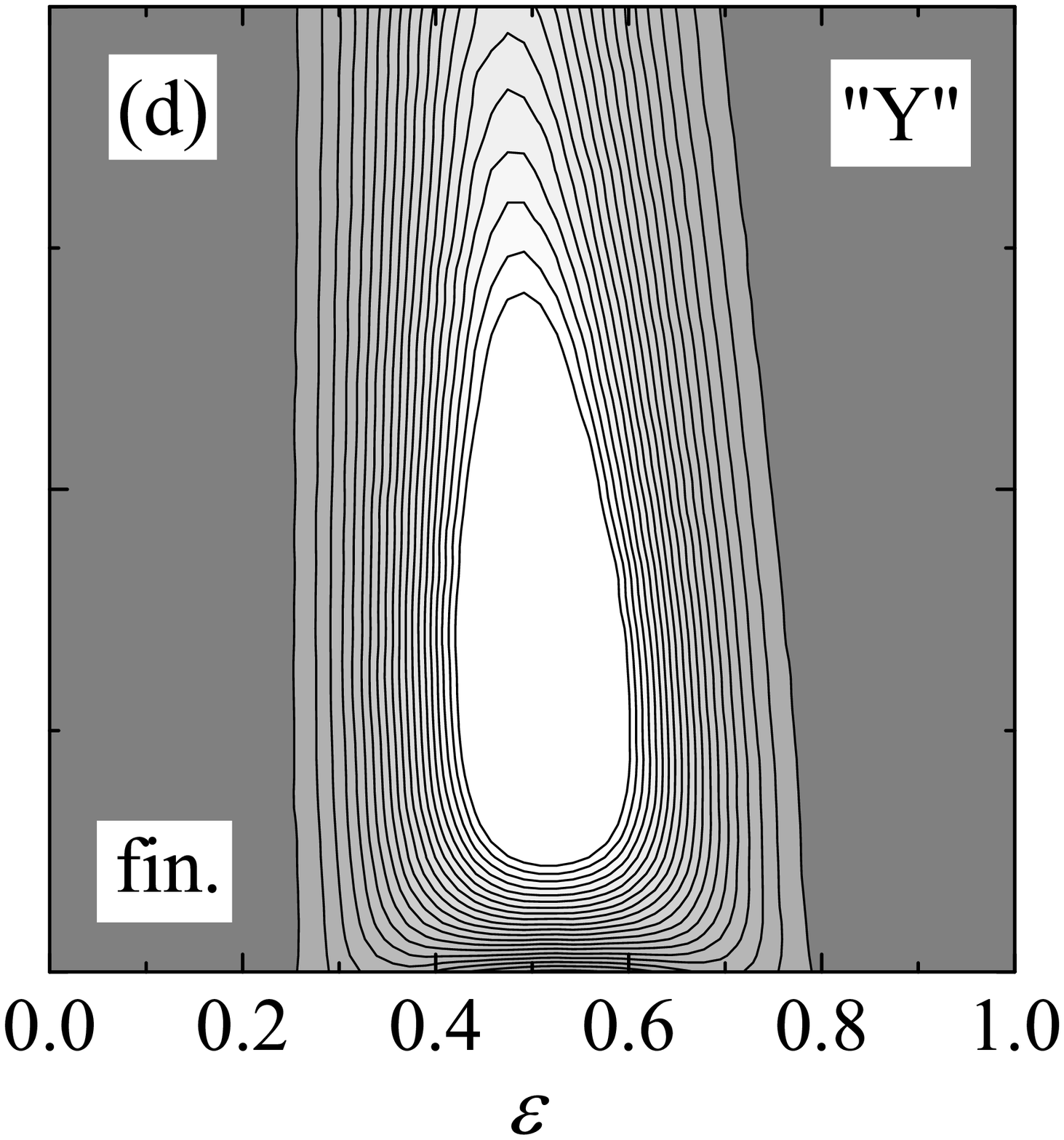}
\end{tabular}
\caption{Contour maps of the distribution density on the kinematical plane
$\{\varepsilon,\cos (\theta_k ) \}$ for $^{19}$Mg in ``T'' (upper row) and ``Y''
(lower row) Jacobi coordinate systems without (left panels, ``init.'') and with
(right panels, ``fin.'') classical extrapolation.}
\label{fig:19mg-complete}
\end{figure}

\begin{figure}[tb]
\begin{tabular}{ll}
\includegraphics[width=0.24\textwidth]{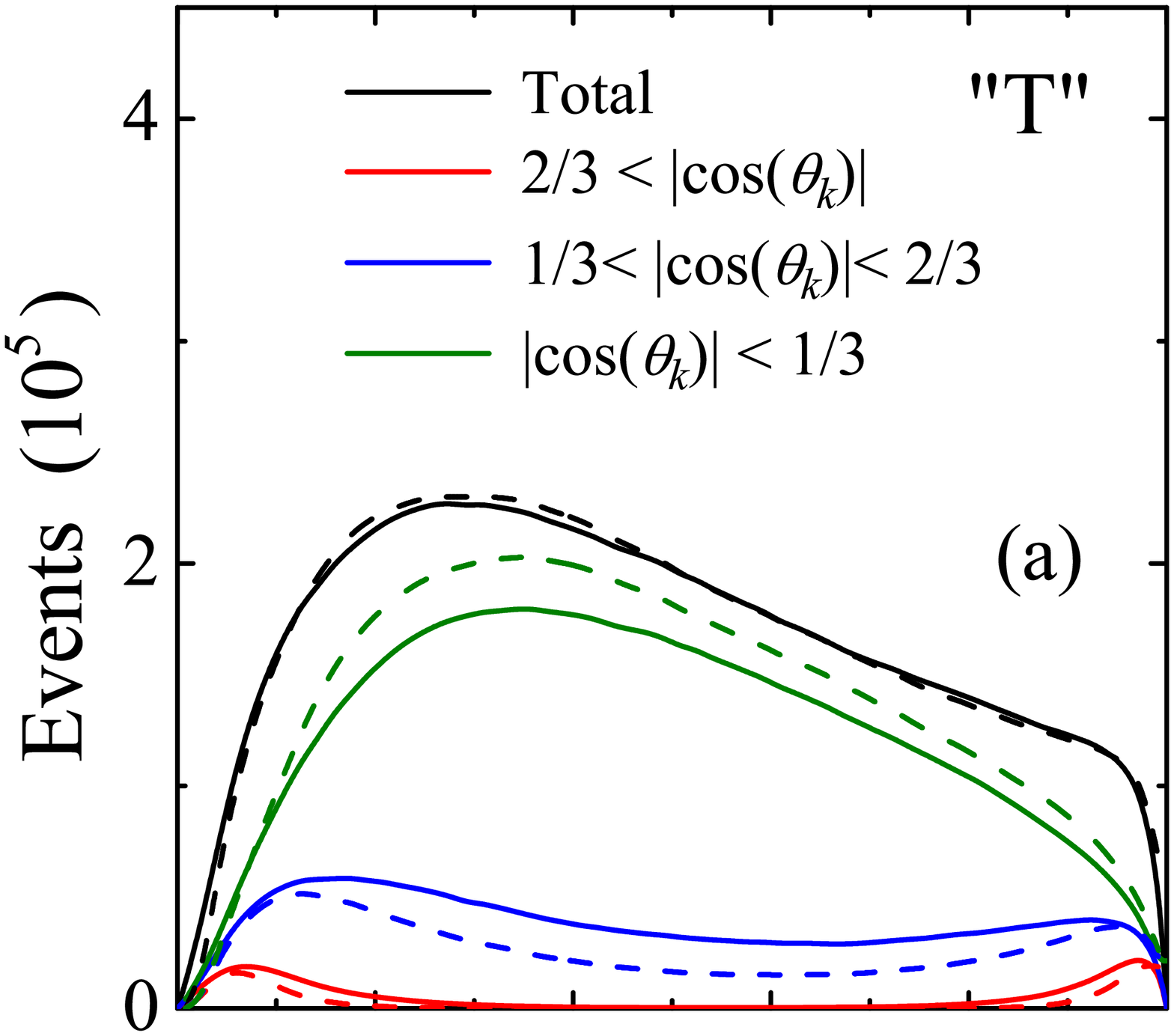}   &
\includegraphics[width=0.214\textwidth]{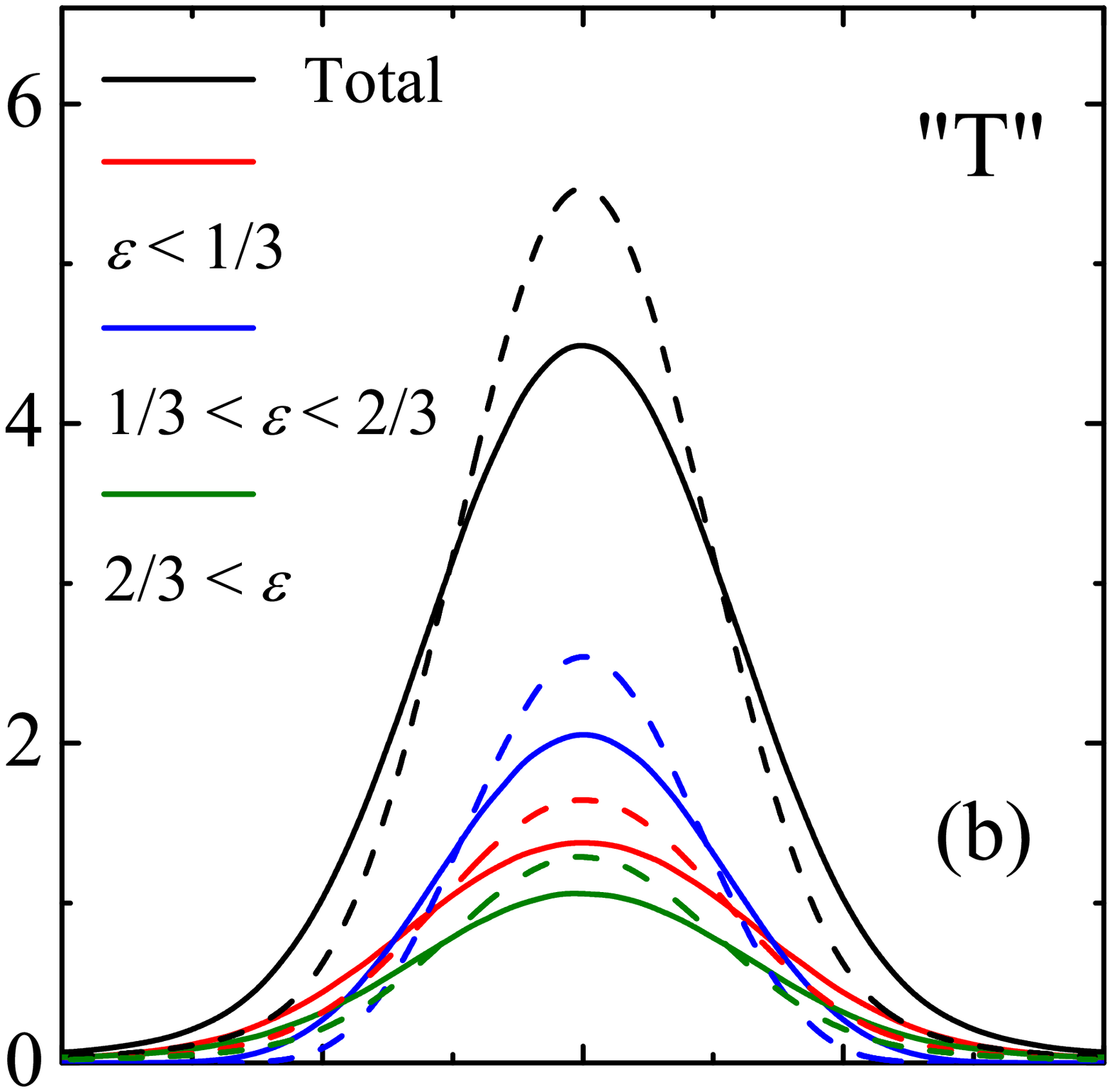} \\
\includegraphics[width=0.25\textwidth]{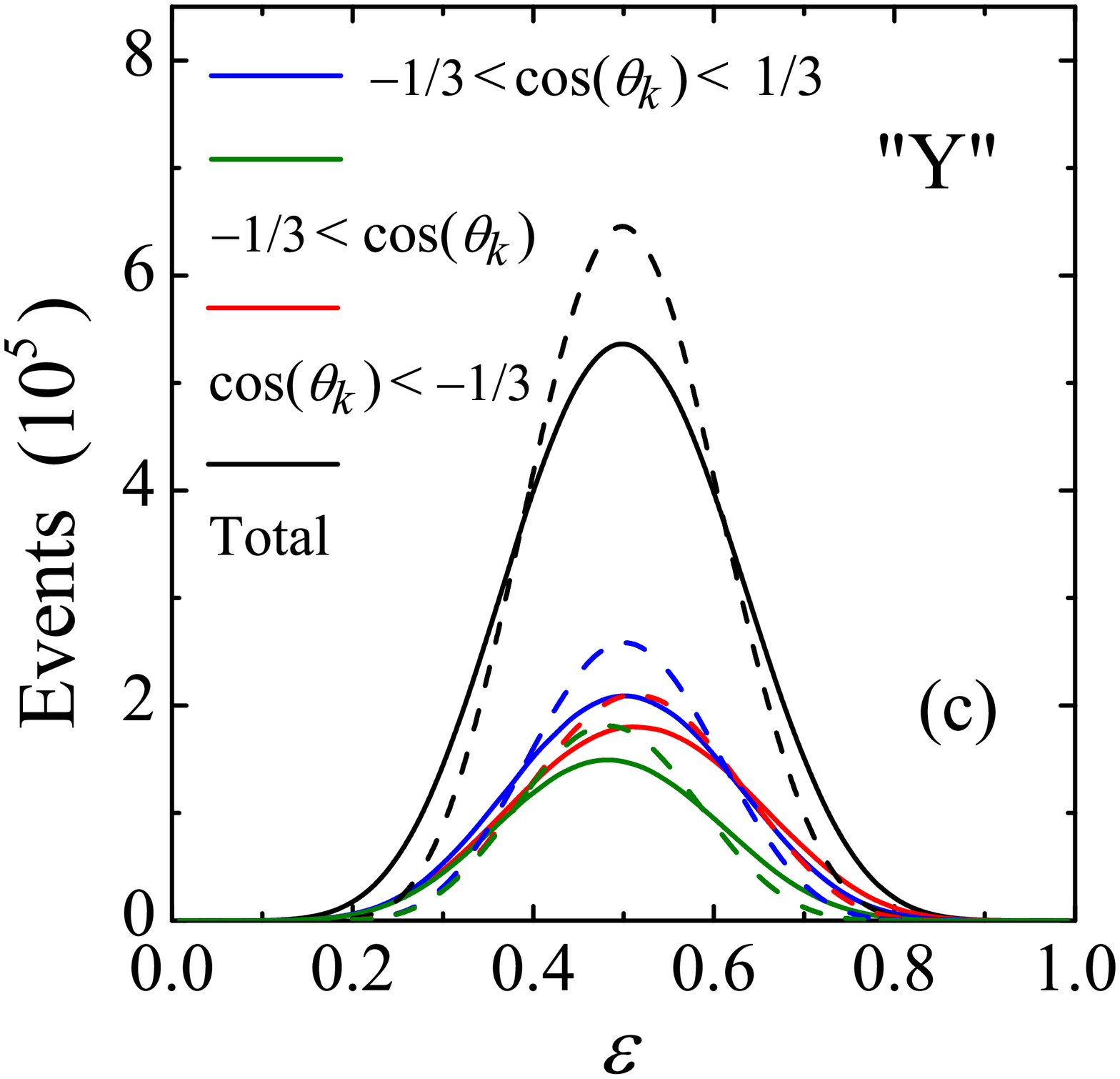}    &
\includegraphics[width=0.226\textwidth]{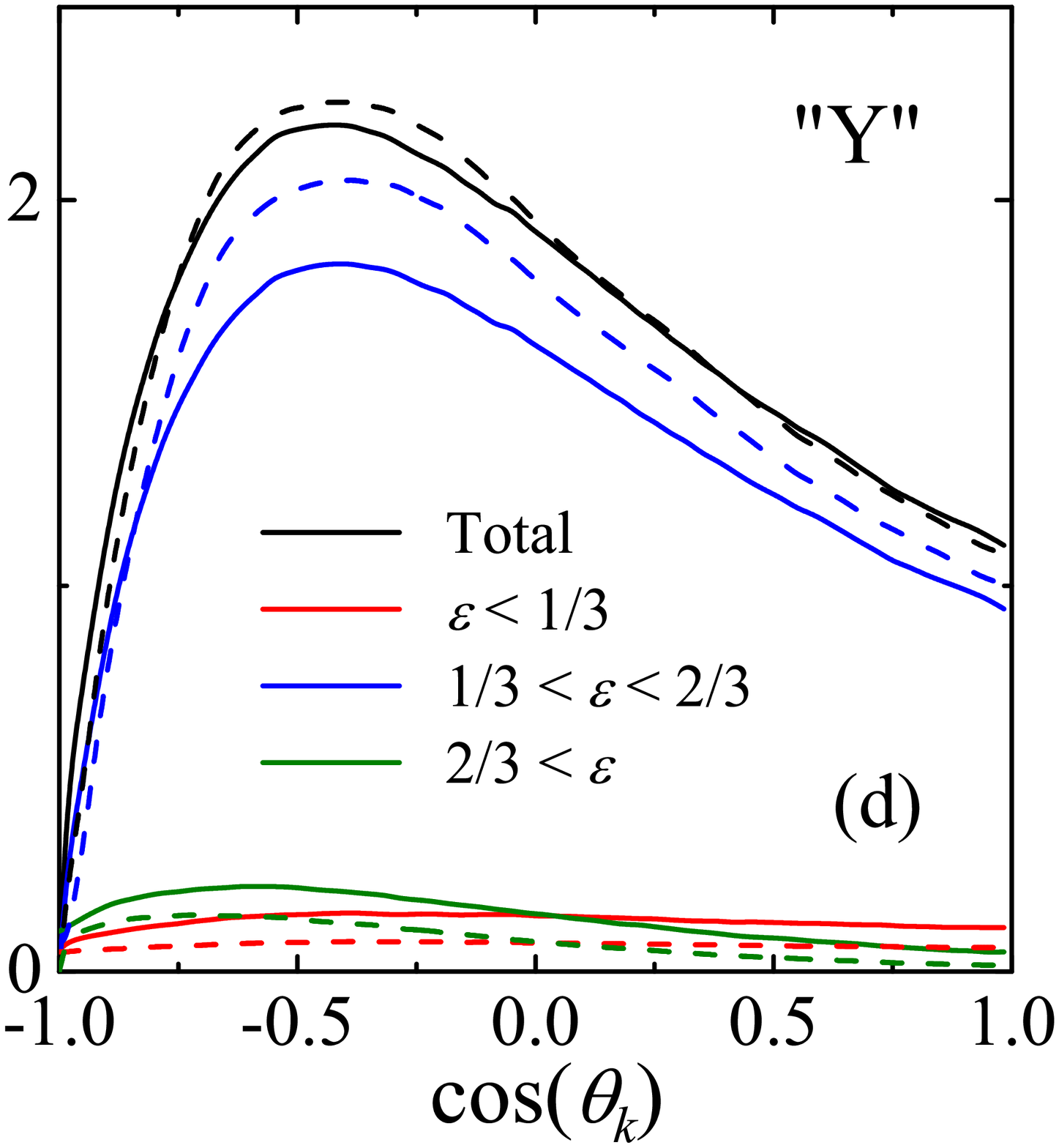}
\end{tabular}
\caption{(Color online) Inclusive energy and angular distributions for $^{19}$Mg
in ``T'' (upper row) and ``Y'' (lower row) Jacobi coordinate systems
without
(solid curves) and with (dashed curves) classical extrapolation. Black lines
show the total distribution and the color coded lines show the inclusive
distributions for certain energy and angular bins (described in the legends).}
\label{fig:19mg-proj}
\end{figure}

An improvement of the momentum distribution due to classical extrapolation is
demonstrated in Fig.\ \ref{fig:19mg-complete} for the complete momentum
distributions and in Fig.\ \ref{fig:19mg-proj} for the inclusive ones. It can be
seen that the angular distributions in the ``T'' and the energy distributions in
the ``Y'' Jacobi systems are the most sensitive to the extrapolation. The effect 
of
the extrapolation on the distributions in certain energy and angular bins can be
very large. The energy distribution in the ``T'' system is only slightly
modified by the CE, but it is interesting to note that for very small
$\varepsilon$ values (where $p$-$p$ Coulomb interaction is expected to be most
active) the extrapolated distribution is visibly suppressed.

Unfortunately the available experimental data on the momentum distributions in
$^{19}$Mg \cite{muk07} do not provide complete distributions, but provide
distributions projected on a plane (perpendicular to the incident beam axis).
Such distributions integrated over one variable have lost some information and
can be more complicated to interpret.


\subsection{Decay of the $^{45}$Fe}

\begin{figure}[tb]
\includegraphics[width=0.42\textwidth]{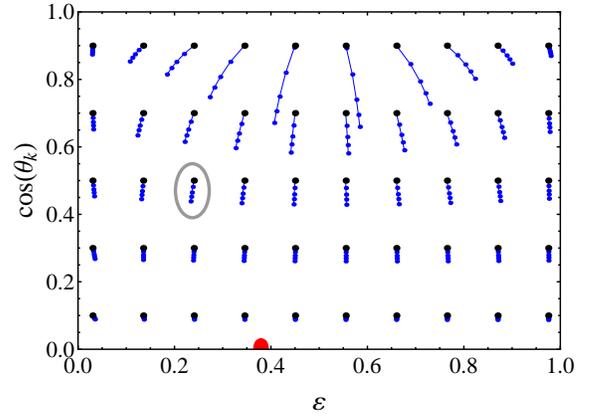}
\caption{(Color online) Classical trajectories on the kinematical
plane $\{\varepsilon, \cos (\theta_k)\}$ for $^{45}$Fe in the Jacobi ``T''
system, $E_T=1.154$ MeV. Starting points (larger black dots) correspond to
$\rho_{\max}=1000$ fm. Dots in the curves correspond to $\rho_{\text{ext}}$
equal 1400, 2200, 4000, and $10^5$ fm.  The red dot at the axis
$\cos(\theta_k)=0$ corresponds to a stationary point, see the discussion in
Sec.\ \ref{sec:ss-solutions}.}
\label{fig:fe-traj}
\end{figure}

The  $^{45}$Fe nucleus is the heaviest $2p$ emitter studied so far and the
effect of the CE is the largest, see Fig.\ \ref{fig:fe-traj}.

\begin{figure}
\centerline{
\includegraphics[width=0.248\textwidth]{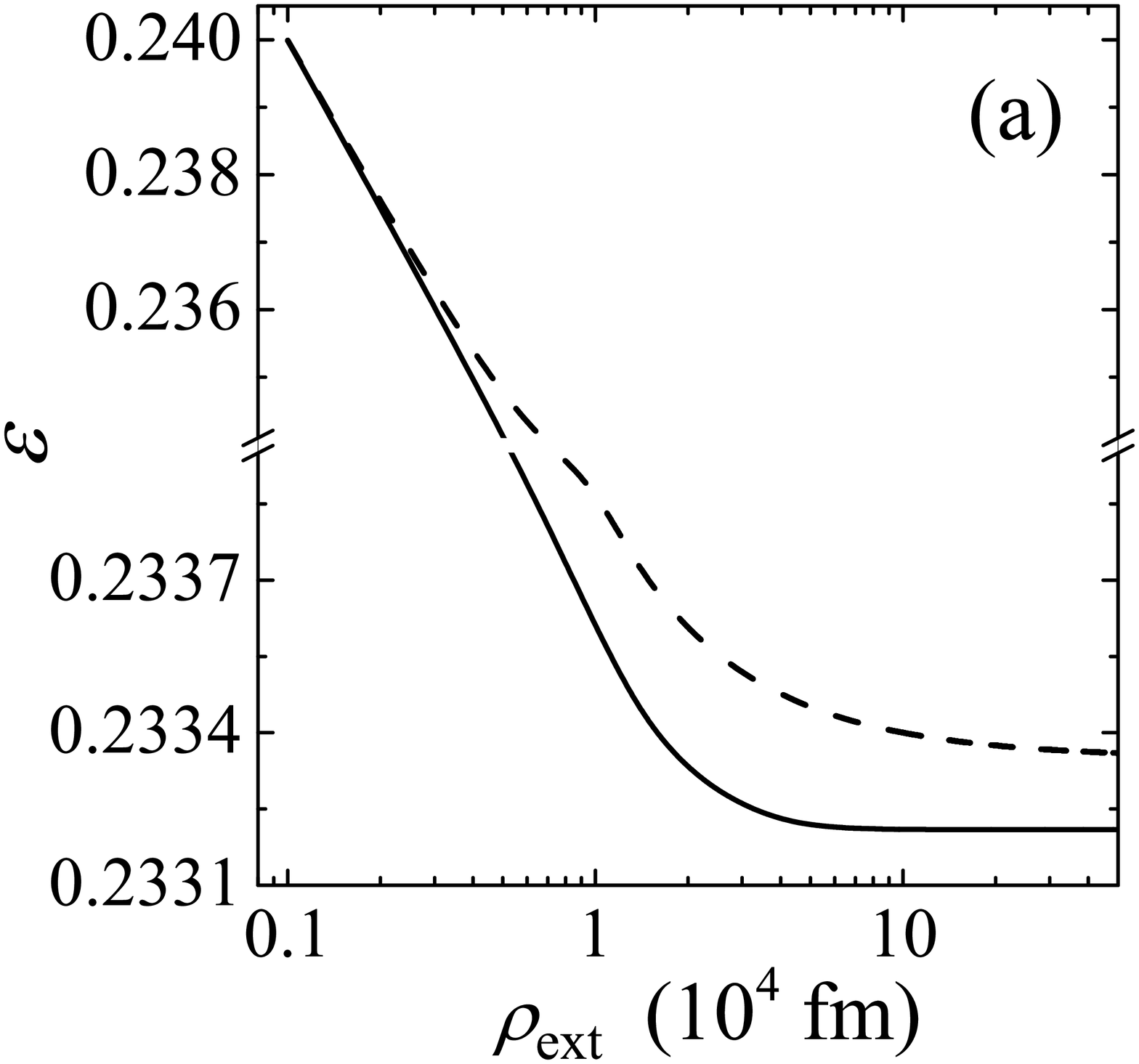}
\includegraphics[width=0.242\textwidth]{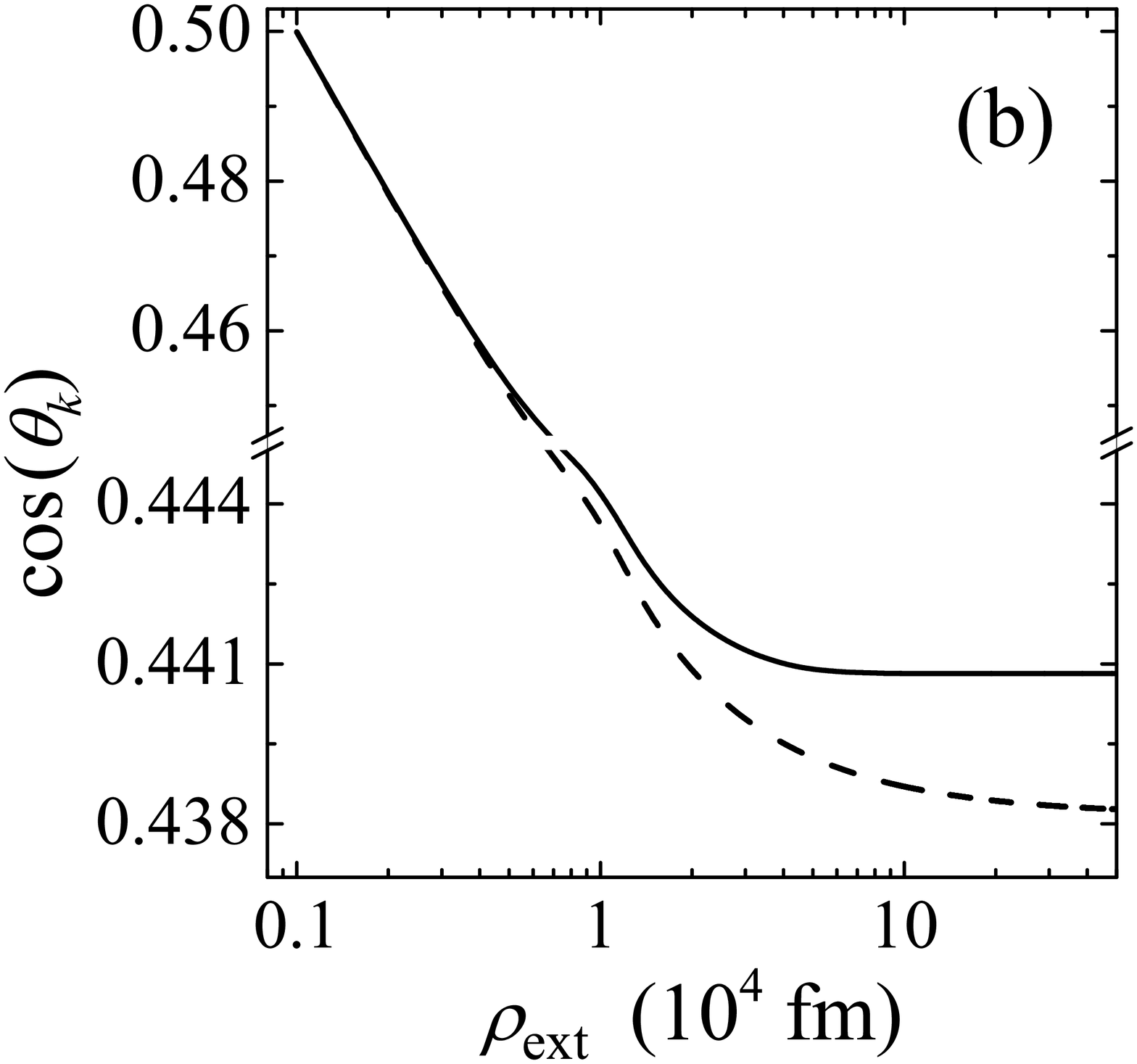}
}
\caption{Radial stabilization of the values $\varepsilon$ [panel (a)] and
$\cos(\theta_k)$  [panel (b)] with $\rho$ in the case of screened Coulomb
potential (solid curves) and in the case of nuclear Coulomb potential only
(dashed curves) for one selected trajectory in $^{45}$Fe (see Fig.\
\ref{fig:fe-traj}). $E_T=1.154$ MeV, $\rho_{\max}=1000$ fm.}
\label{fig:fe-traj-scr}
\end{figure}

Radial stabilization of the values $\varepsilon$ and $\cos(\theta_k)$ for one
selected trajectory is demonstrated in Fig.\
\ref{fig:fe-traj-scr} (this trajectory is shown in the gray ellipse in Fig.\
\ref{fig:fe-traj}). The trajectories are well ``converged'' by about $(3-4)
\times 10^4$ fm but some drift continues up to much larger $\rho$ values. In
real experimental situations, this slow drift can be suppressed by electron
screening which is discussed separately in Sec.\ \ref{sec:scr}.

\begin{figure}[tb]
\begin{tabular}{cc}
\includegraphics[width=0.251\textwidth]{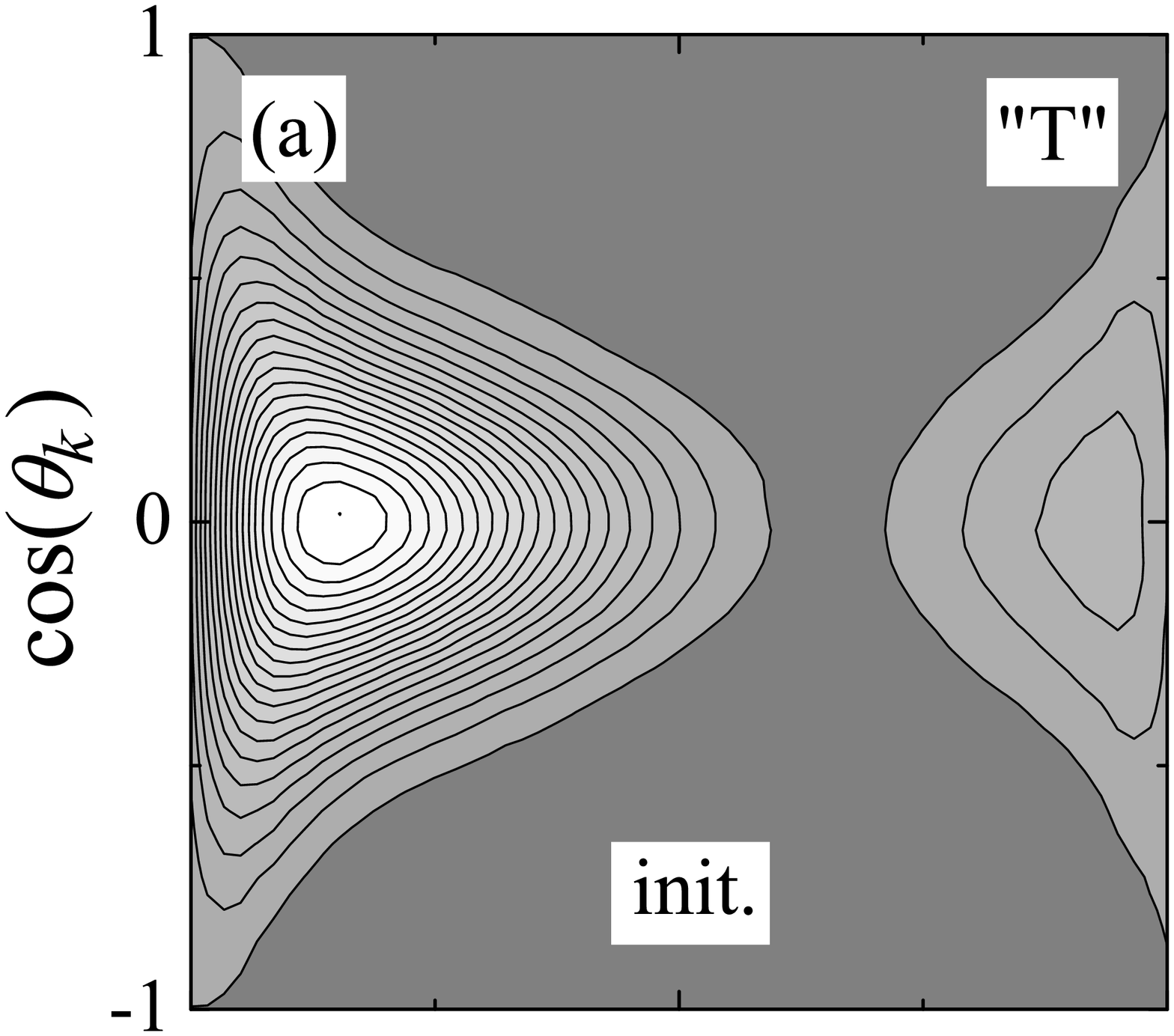}   &
\includegraphics[width=0.22\textwidth]{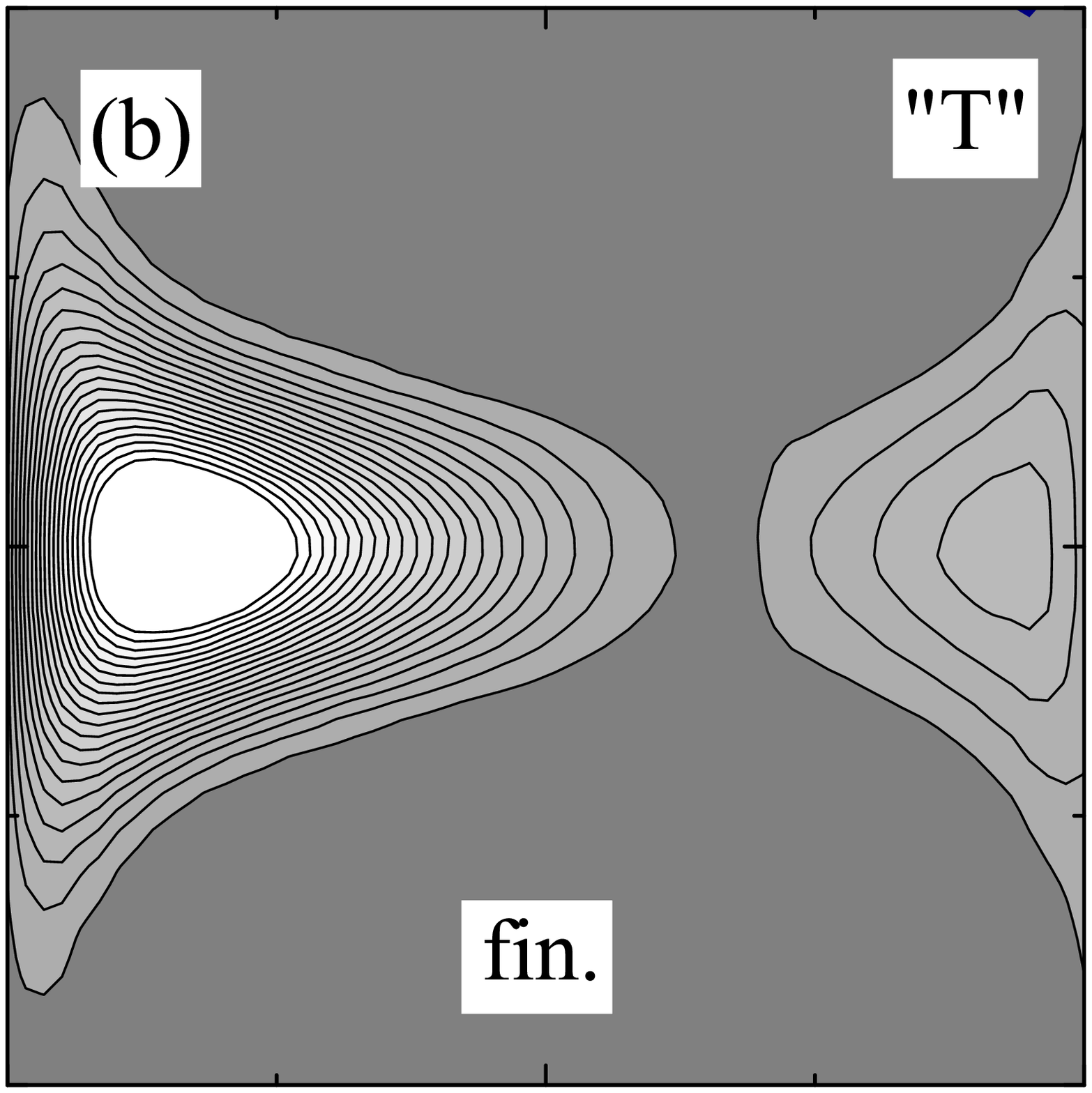} \\
\includegraphics[width=0.251\textwidth]{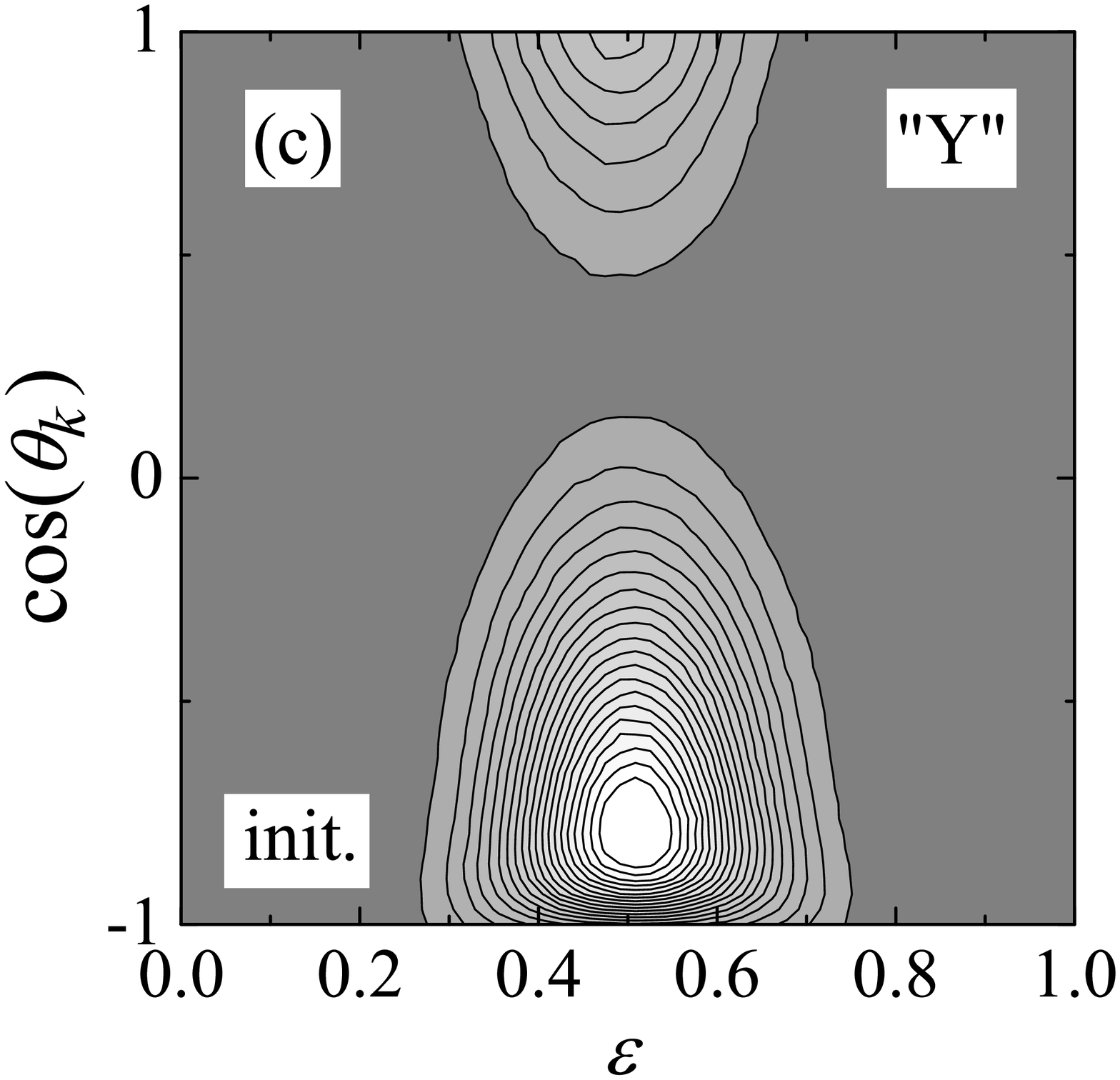}    &
\includegraphics[width=0.22\textwidth]{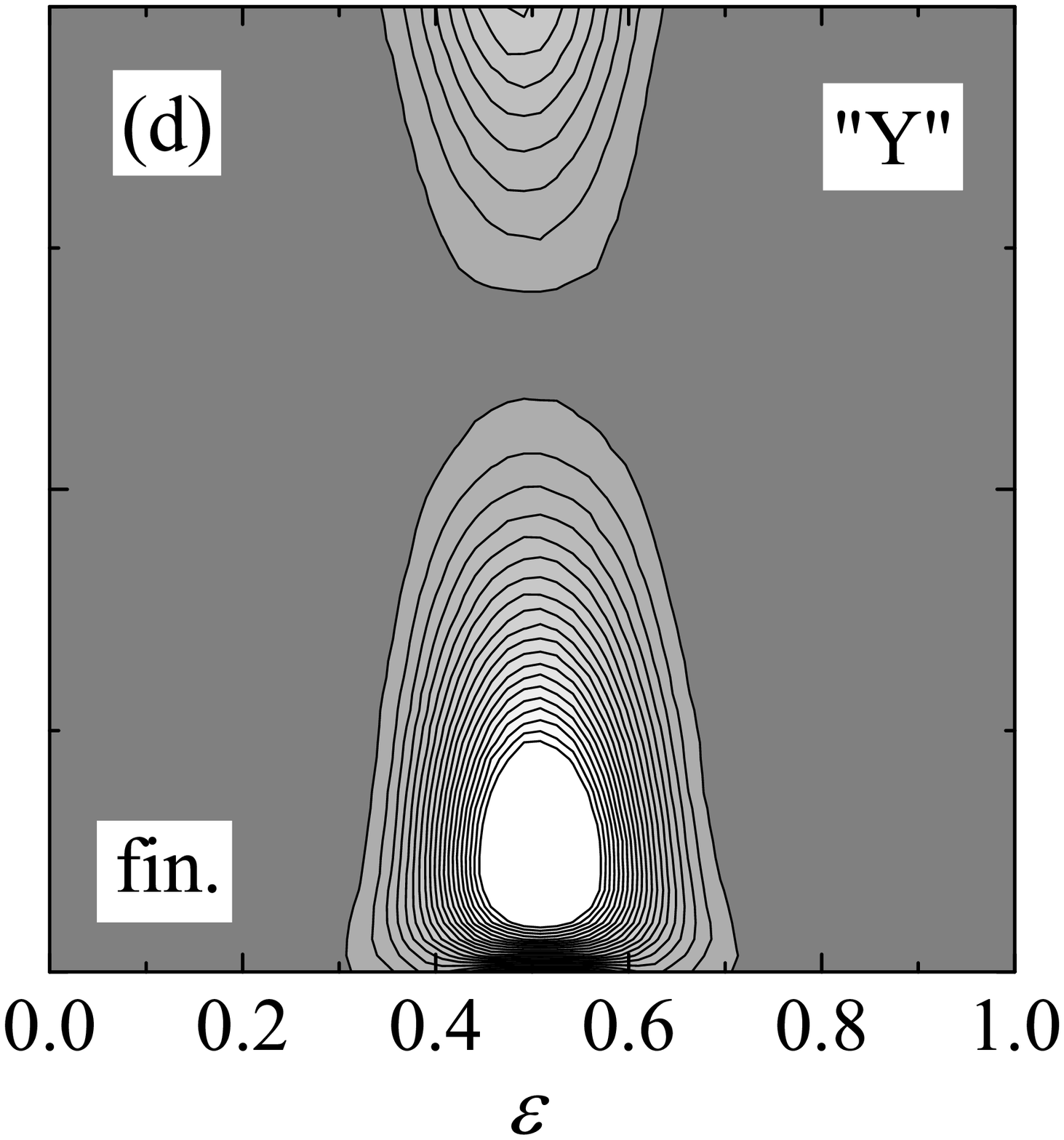}
\end{tabular}
\caption{Contour maps of the distribution density on the
kinematical plane $\{\varepsilon,\cos (\theta_k) \}$ for $^{45}$Fe in ``T''
(upper row) and ``Y'' (lower row) Jacobi coordinate systems without (left
panels, ``init.'') and with (right panels, ``fin.'') classical extrapolation.}
\label{fig:45fe-complete}
\end{figure}

\begin{figure}[tb]
\begin{tabular}{ll}
\includegraphics[width=0.24\textwidth]{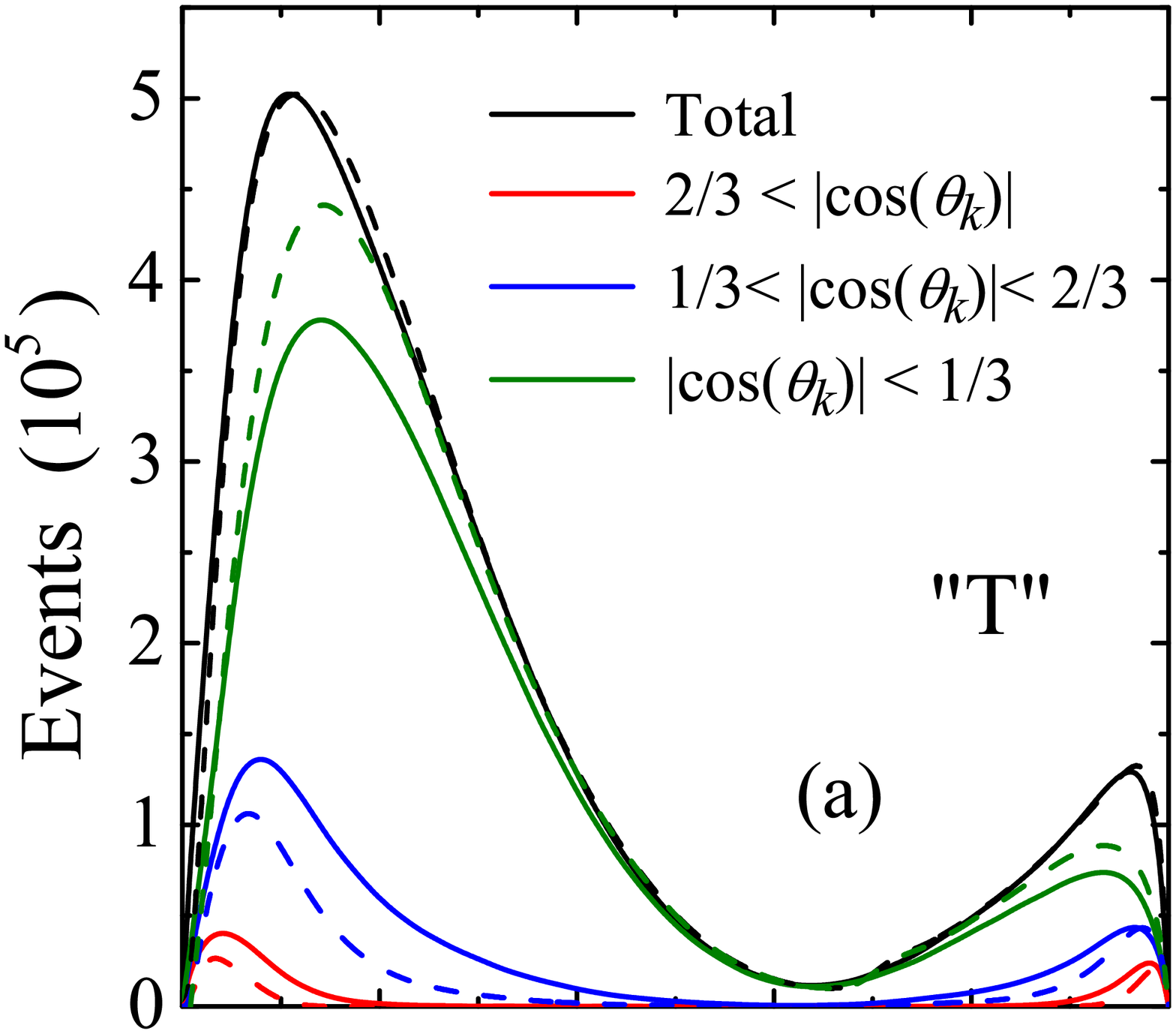}   &
\includegraphics[width=0.214\textwidth]{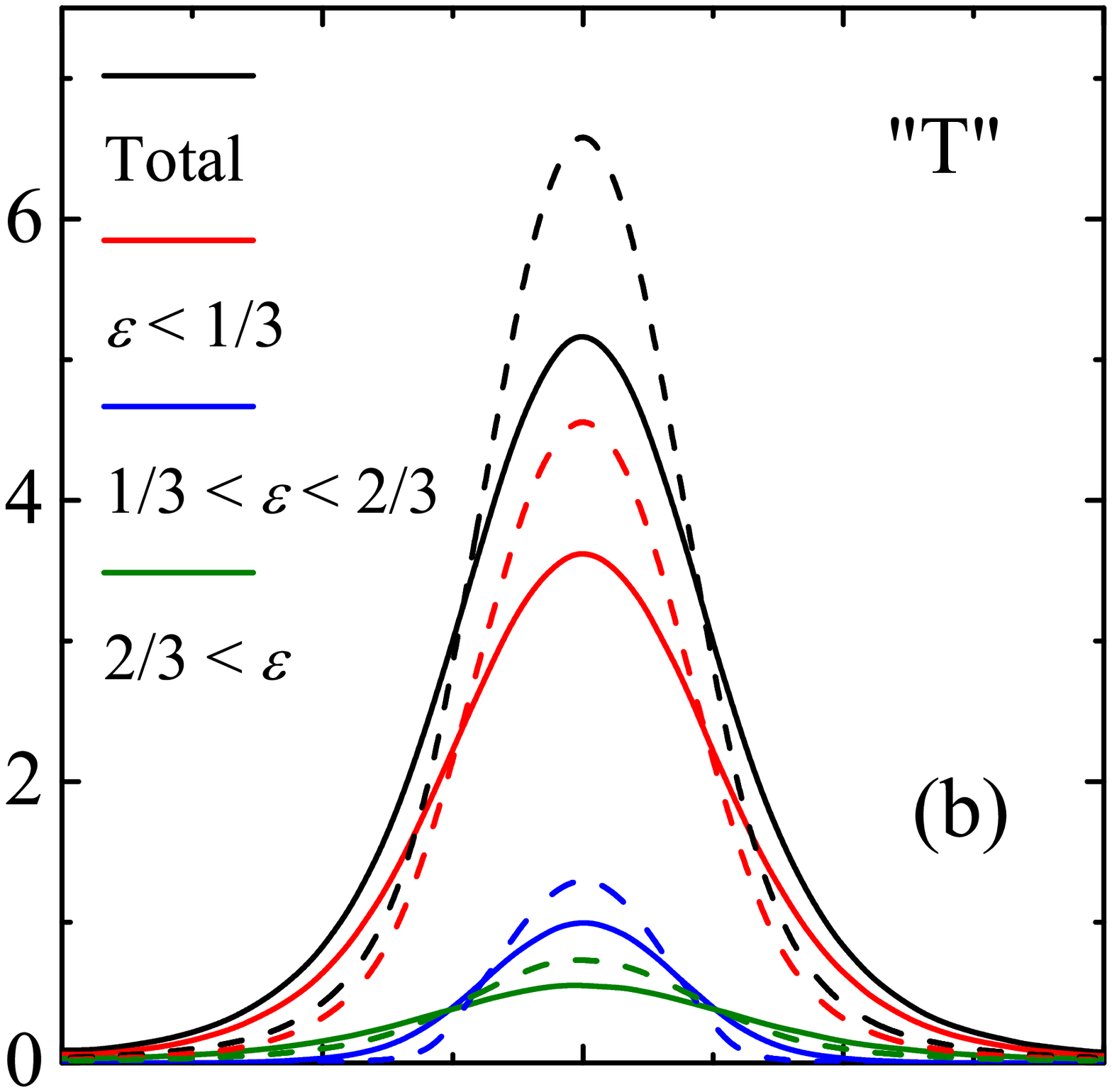} \\
\includegraphics[width=0.25\textwidth]{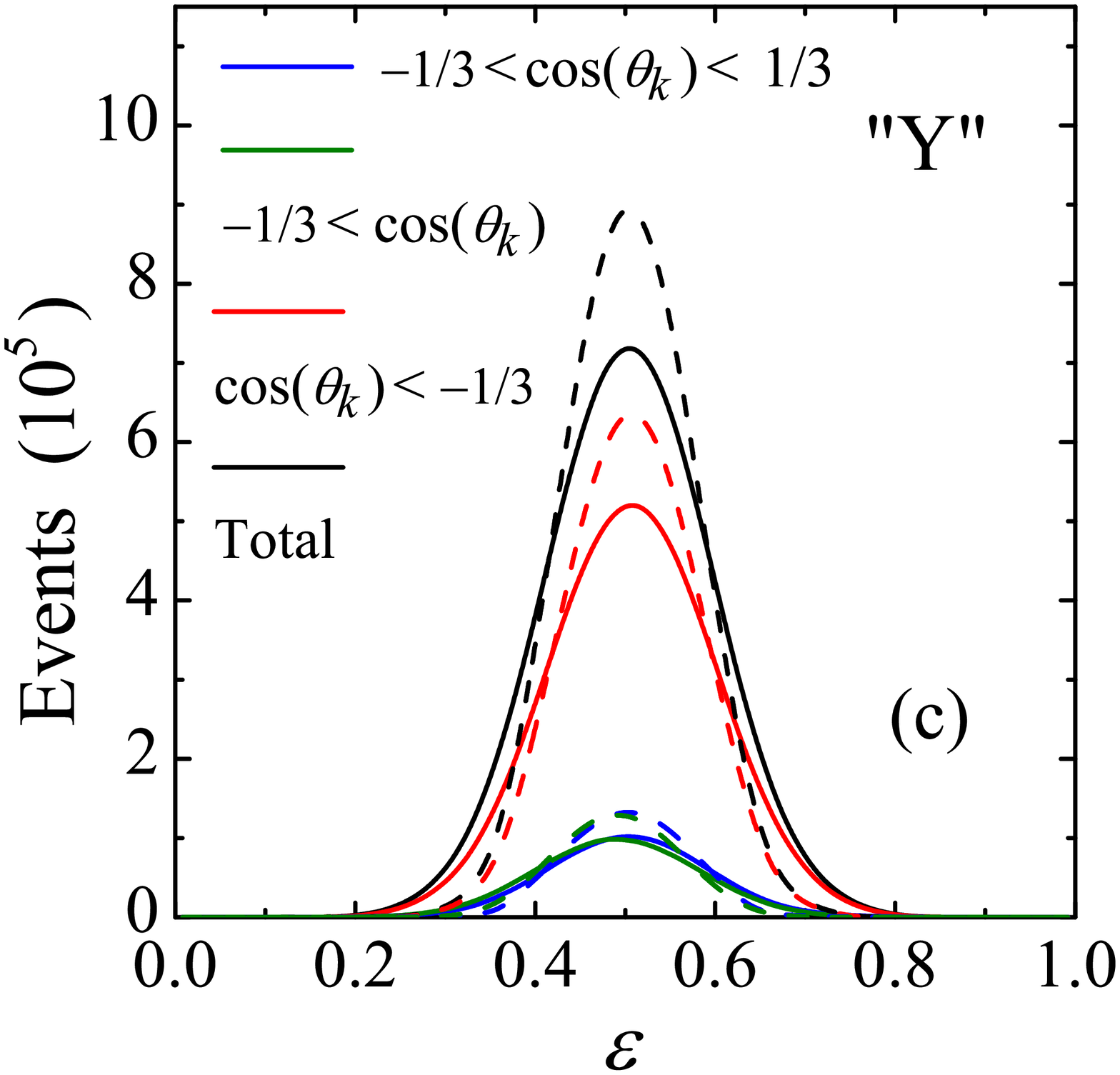}    &
\includegraphics[width=0.226\textwidth]{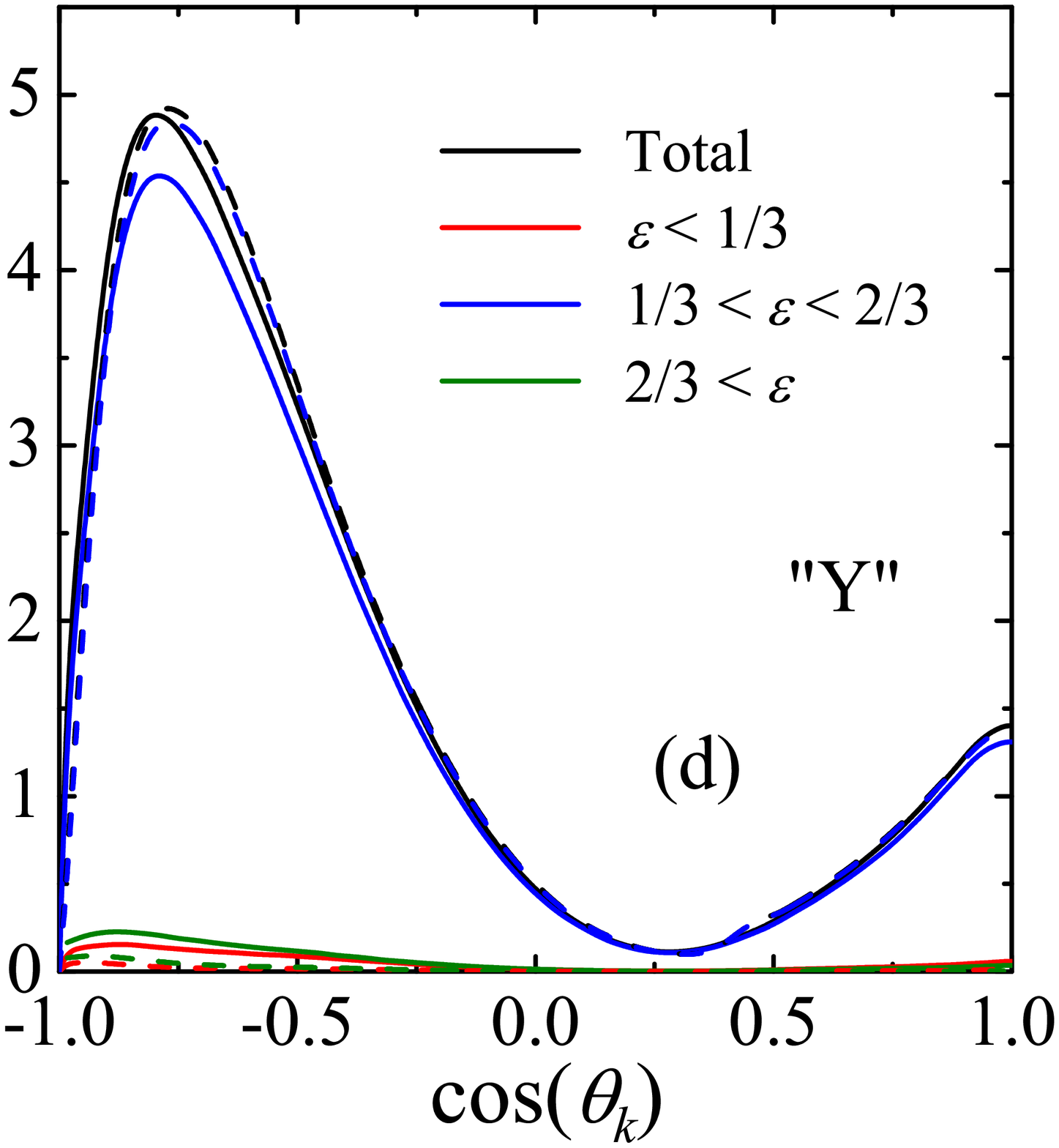}
\end{tabular}
\caption{(Color online) Inclusive energy and angular distributions for $^{45}$Fe
in ``T'' (left column) and ``Y'' (right column) Jacobi coordinate systems
without (solid curves) and with (dashed curves) classical extrapolation. Black
lines show the total distribution and the color coded lines show the inclusive
distributions for certain energy and angular bins (described in the legends).}
\label{fig:45fe-proj}
\end{figure}

The improvement of the momentum distributions due to the classical extrapolation 
for
$^{45}$Fe is demonstrated in Fig.\ \ref{fig:45fe-complete} for the  complete
momentum distributions and in Fig.\ \ref{fig:45fe-proj} for the inclusive ones.
The most impressive modifications are for the $\varepsilon$ distribution in the
``Y'' system and for the  $\cos(\theta_k)$ distribution in the ``T'' system. As
far as these distributions have bell shapes, centered at (or close to) the 
center
of the kinematical range, we can characterize them in terms of the full width at
 half maximum (FWHM). Classical extrapolation decreases this value by
about $30\%$ for $\cos(\theta_k)$ in ``T'' system and by about $10\%$ for
$\varepsilon$ in ``Y'' system. This effect is sufficiently large to be
already observable at the current level of the experimental precision.

\begin{figure}[tb]
\includegraphics[width=0.235\textwidth]{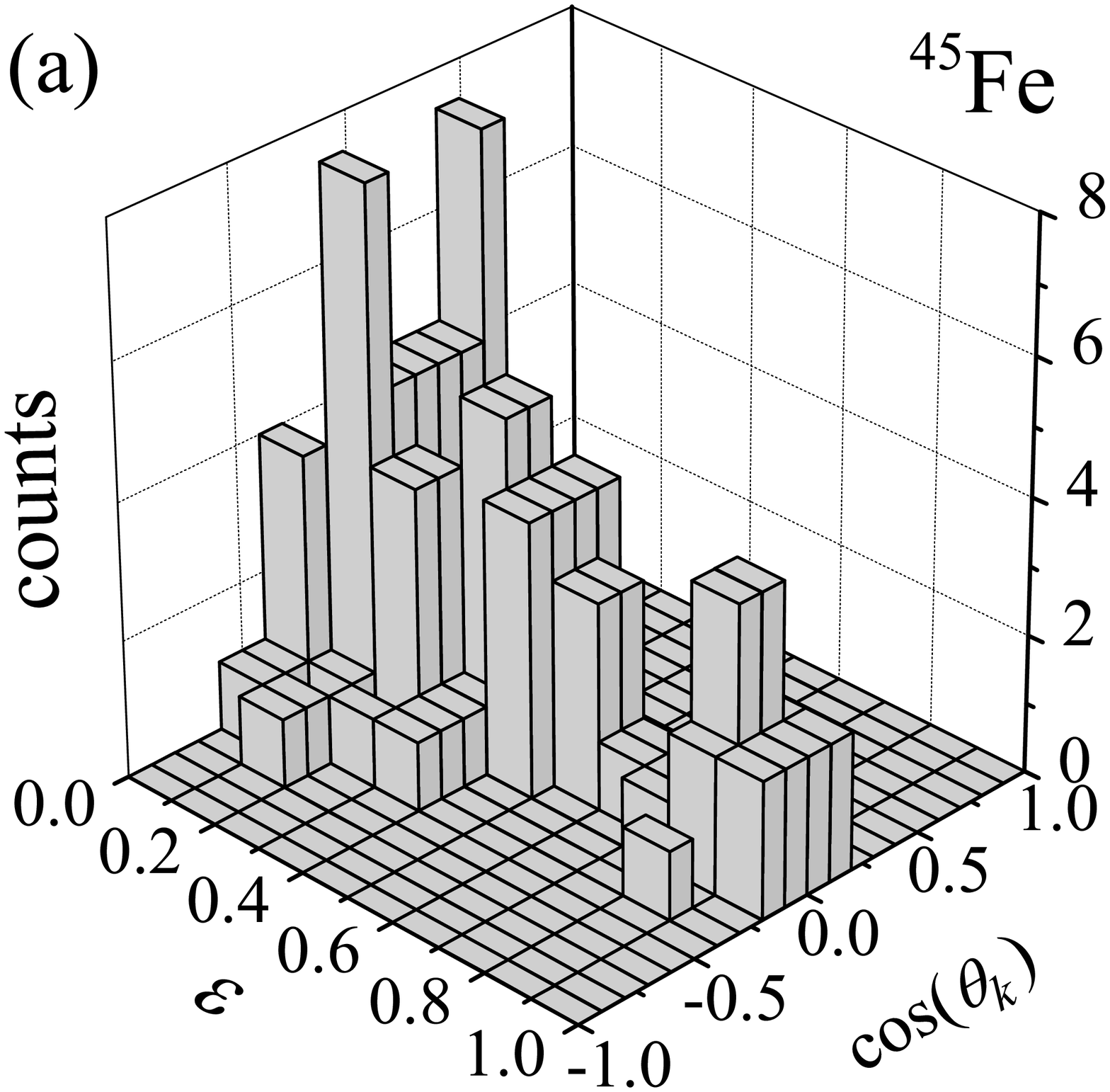}
\includegraphics[width=0.234\textwidth]{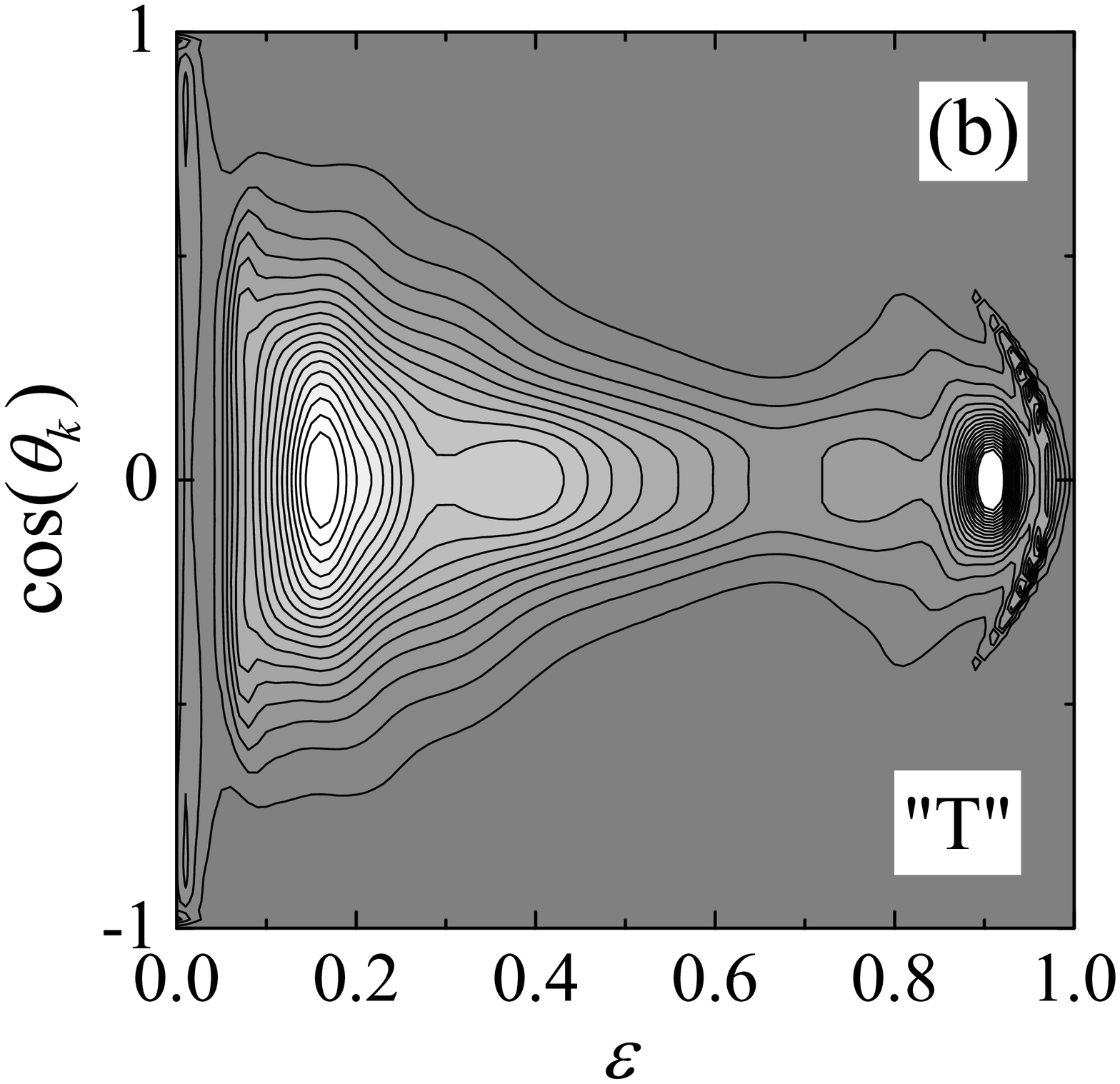}
\caption{Experimental distribution density in $^{45}$Fe in ``T'' Jacobi system.
Original distribution from \cite{mie07} is shown in the panel (a) as a
histogram. ``Smooth'' version of this distribution taking into account
experimental errors is shown in the panel (b) as a contour plot.}
\label{fig:45fe-exp}
\end{figure}

The experimental distribution for $^{45}$Fe \cite{mie07} has quite low 
statistics
(150 events) and therefore it is far from being smooth, see Fig.\
\ref{fig:45fe-exp} (a). To make a visual comparison with theoretical
calculations possible, we produce a ``smooth'' representation of this data based
on the experimental uncertainties. The raw
experimental data measured in \cite{mie07} by an optical time projection chamber
consists of the energies and the polar angles of the two protons and the 
azimuthal
angle between the projections of the two protons' momenta on the cathode plane 
of the
chamber. Each parameter for each event has a value (and its uncertainty) defined
individually by a complex iterative fitting procedure. Instead of each event, we
generate an event distribution based on the stochastic Gaussian variation of 
each
parameter within its uncertainty range. So instead of one point in the kinematic
space we get a kind of a ``probability cloud''. The result of this procedure is
shown in Fig.\ \ref{fig:45fe-exp} (b). This procedure is not a cure for small
statistics, but for small statistics and large experimental uncertainties we
think it is a preferable presentation as it incorporates information about the
distortions caused by the measurement procedure in a consistent and visible way.

\begin{figure}[tb]
\centerline{\includegraphics[width=0.4\textwidth]{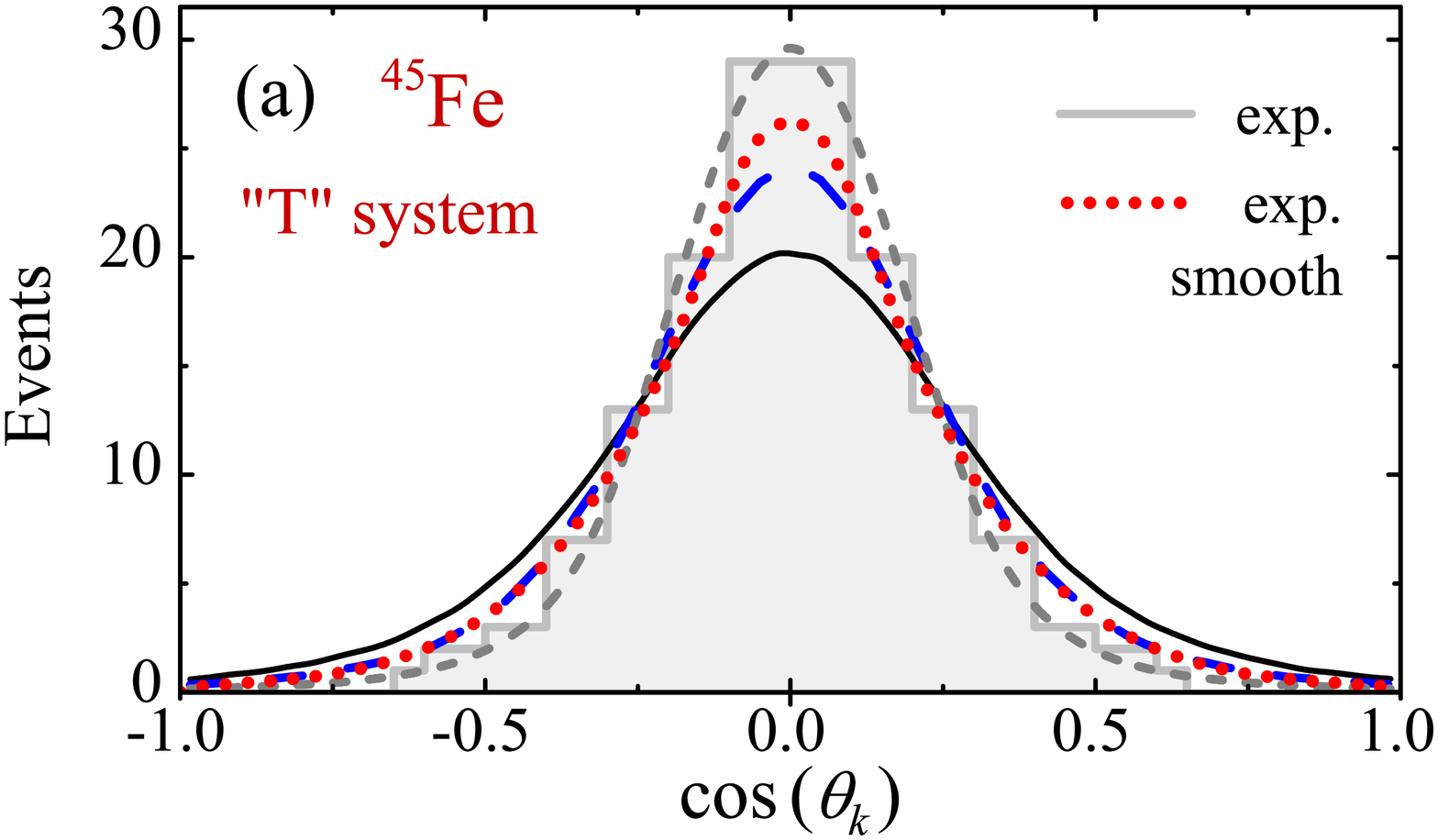}}
\centerline{\includegraphics[width=0.4\textwidth]{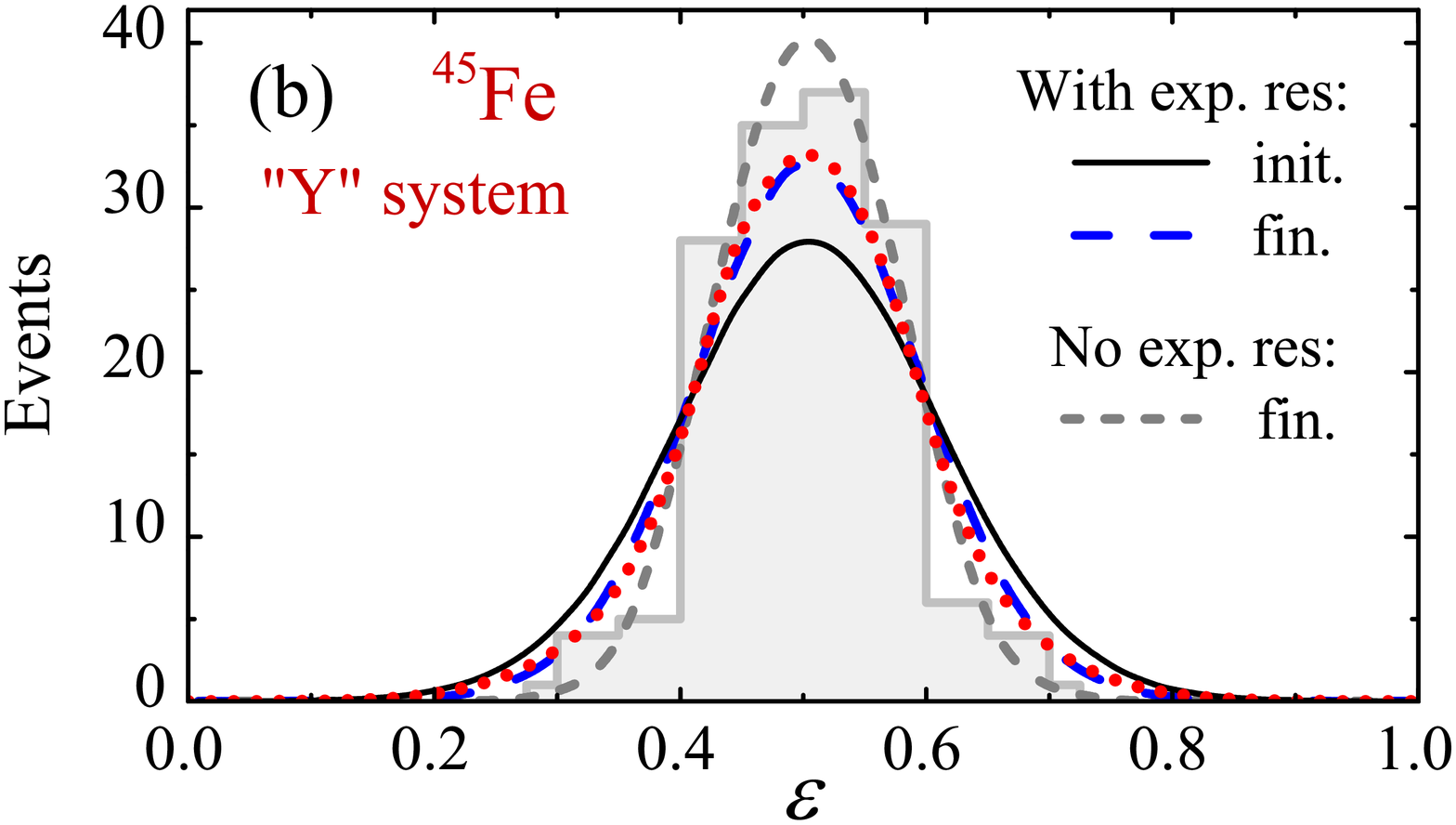}}
\caption{(Color online) Inclusive angular (a) and energy (b) distributions for
$^{45}$Fe in ``T'' and ``Y'' Jacobi coordinate systems without (``init.'') and
with (``fin.'') classical extrapolation compared to the experimental data. The
results with experimental resolution of \cite{mie07} and without
it are explained in legends, which are the same for both panels.}
\label{fig:45fe-exp-proj}
\end{figure}

The experimental data are compared with inclusive theoretical distributions
sensitive to the classical extrapolation in Fig.\ \ref{fig:45fe-exp-proj}. In
this plot, theoretical results were treated by the procedure which is maximally
close to the experimental treatment of the data: (i) for the ``theoretical
event'' the nearest experimental event in the space of parameters
$\{E_{p1},E_{p2},\theta_1,\theta_2,|\phi_2-\phi_1|\}$ was defined, (ii)
spherical coordinates for protons from the ``theoretical event''  were
distributed according to the errors of the nearest experimental event, (iii) the
momentum of the core was reconstructed and the total energy of the ``distorted''
theoretical event was renormalized to correspond exactly to the experimental
one, and (iv) a new location in the kinematical plane
$\{\varepsilon,\cos(\theta_k)\}$ was defined. The effect of the experimental
resolution is a roughly $25\%$ increase of FWHM for the $\varepsilon$ 
distribution and
an $18\%$ increase of FWHM for the $\cos(\theta_k)$ distribution (see Fig.\
\ref{fig:45fe-exp-proj}). It can also be seen in Fig.\ \ref{fig:45fe-exp-proj}
that the theoretical results with classical extrapolation are in quantitative
agreement with the experiment, while without CE they are not completely
consistent with the data. So, we have appreciable experimental evidence that the
long-range treatment of the momentum distributions (namely CE) is necessary  for
heavy $2p$ emitters.

\section{Discussion}
\label{sec:disc}



\subsection{Classical motion}
\label{sec:class}


It is important to note that large hyperradii are used to start the classical
extrapolation procedure. Specifically for true $2p$ decay with such large
hyperradii, practically the whole WF resides is in the classically allowed 
region
(probability to find the system in the classically forbidden region is very
small). For example, for the $^{45}$Fe calculation with hyperradius
$\rho_{\max}=1000$ fm and MC generation of $10^7$ events it is typical that not
a single event is generated which is situated in the classically forbidden
region. This fact confirms the validity of the choice of a hypersphere as the
surface at which the switching from quantum-mechanical to classical methods is
performed.


\subsection{Electron screening}
\label{sec:scr}


The discussion of the $^{45}$Fe case can provide an illustrative example here. 
So
far, the decay process of $^{45}$Fe with the half-life of 2.6 ms \cite{mie07} 
was
measured in gas (or solid state) detectors. This means that at the  moment of 
decay,
$^{45}$Fe has completely recovered electron shell. The Bohr radius for
$^{45}$Fe is
\begin{equation}
a_0=\frac{1}{m_e \alpha Z}=2035\; \mbox{fm}\,,
\label{eq:bohr-rad}
\end{equation}
where $Z=\sum_i Z_i$ is total charge of $^{45}$Fe.  Therefore, we can expect
that the screening effect of the inner most electrons becomes observable at
about 2000 fm. Classical trajectories for $^{45}$Fe in kinematical space are
well stabilized by $10^5$ fm, but there is a minor drift up to much larger
distances. It is clear that some effect of the electron screening on the
momentum distributions can be expected.

The binding energy of all electrons estimated as independent particles is
$\sum_i m_e (Z/2n_i \alpha)^2$ ($n_i$ is a principal quantum number of the
shell), which gives 52.3 and 47.8 keV for $^{45}$Fe and $^{43}$Cr,
respectively. So, when $^{45}$Fe emits two protons at least two electrons
should be ejected carrying away 4.5 keV of energy. The estimated velocities of
protons with energies around 0.5 MeV and electrons with energies around 1 keV
are 0.033 and 0.063. These velocities are comparable, which means that the $2p$
decay of atomic $^{45}$Fe would be accompanied by a strong reconstruction of
atomic structures having the same timescale.  It is reasonable therefore to make
estimates of a screening with the $^{45}$Fe electron density, but only for 24
electrons. This will somehow account for the effect of the electron shell
disintegration during the $2p$ decay of $^{45}$Fe  and provide a nuclear plus
atomic Coulomb potential tending to zero at infinity.

\begin{figure}
\centerline{
\includegraphics[width=0.218\textwidth]{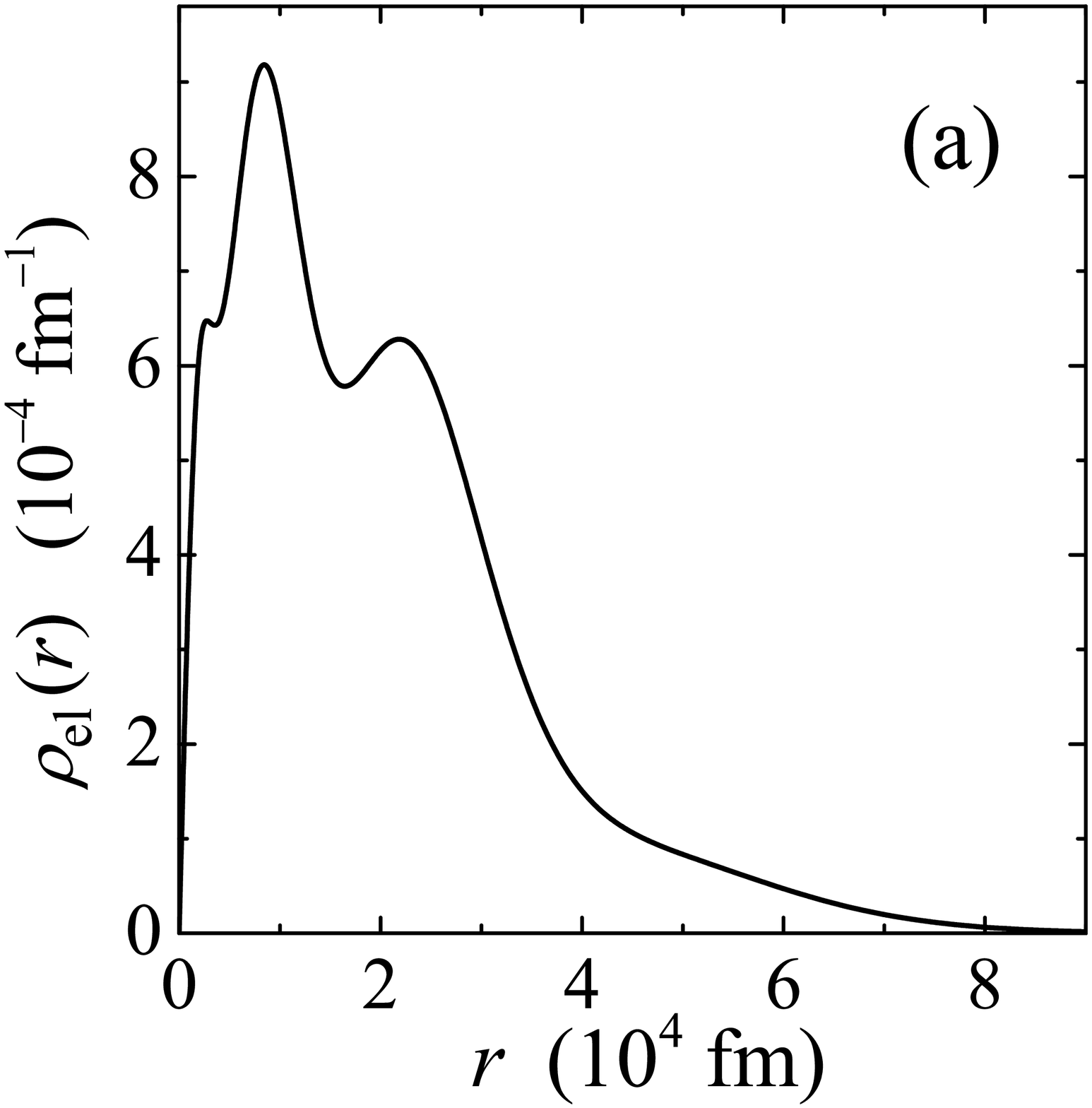}
\includegraphics[width=0.26\textwidth]{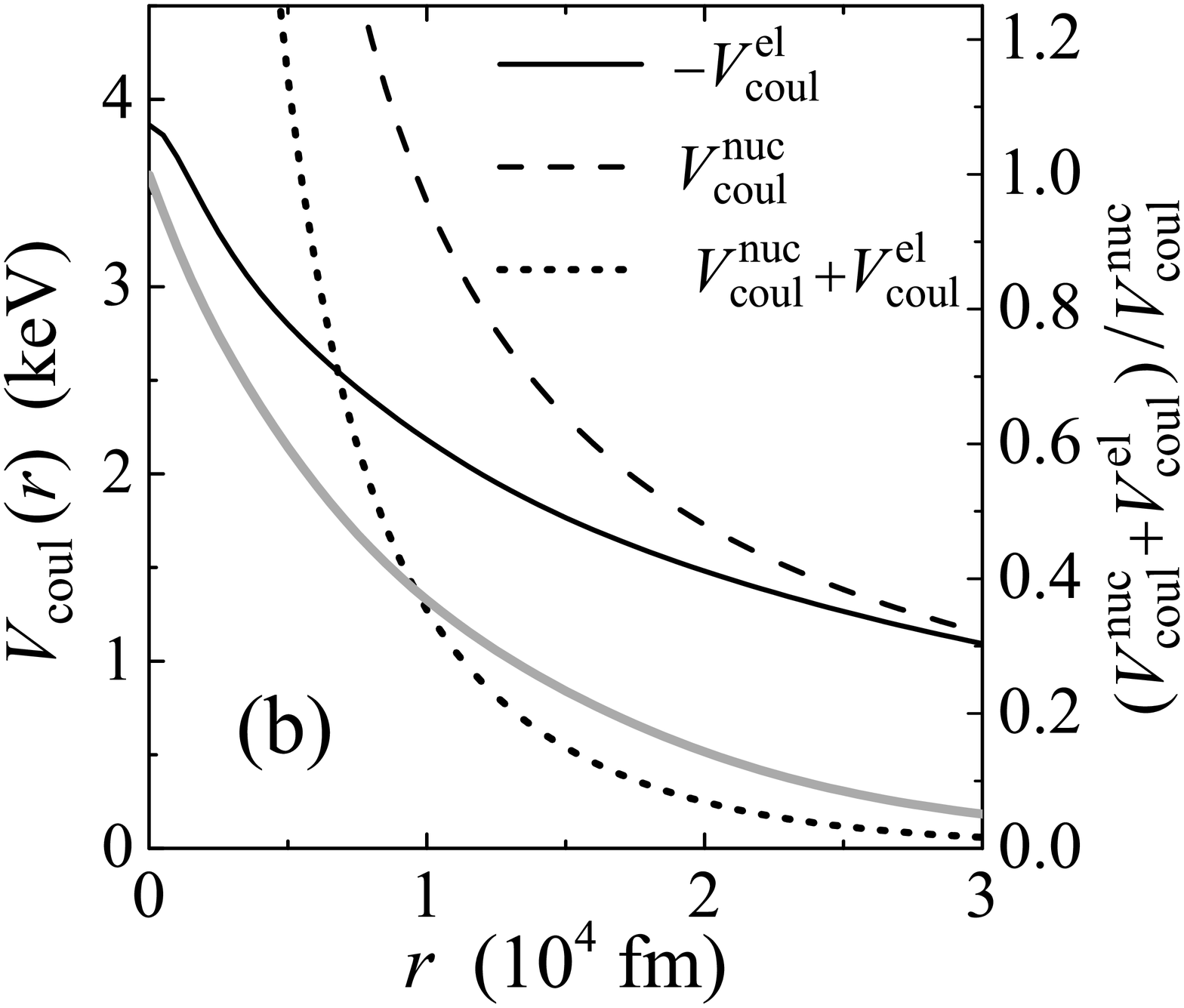}
}
\caption{Panel (a) shows electron density for 24 lowest electron shells in
$^{45}$Fe (normalized for integration over $dr$). Left axis of panel (b) shows
proton potential for Coulomb interaction of nucleus, electron shell, and
their difference (screened potential). The ratio of the screened potential to
the nuclear is shown by gray curve opposite the right axis.}
\label{fig:scr-1}
\end{figure}

The electron density used for the screening calculations and the potentials
obtained are shown in Fig.\ \ref{fig:scr-1}. One can see already that at 2000 
fm, the full
$(V^{\text{nuc}}_{\text{coul}}+ V^{\text{el}}_{\text{coul}})$ Coulomb potential
is noticeably reduced due to the screening compared to nuclear Coulomb potential
(the reduction factor is 0.8). At 7000 fm the reduction factor is 0.5 and it
tends to zero at 30000 fm.

The radial stabilization of the values $\varepsilon$ and $\cos(\theta_k)$ in the
screening case compared to the purely nuclear case is shown in Fig.\
\ref{fig:fe-traj-scr} for one selected trajectory. It can be seen that in the
screening case, the trajectory stabilizes at $\rho \sim (3-4) \times 10^4$ fm. 
In
the purely nuclear case, the minor drift of the trajectory continues to much 
larger
$\rho$ values. The calculations show that in the ``T'' system, the screening
effect is largest for the variable $\cos(\theta_k)$. It is typically at the
level of $0.6 \%$ of the absolute value of this variable and, for
$\rho_{\max}=1000$ fm, it typically accounts for $3-4\%$ of the CE effect. For
an effect which is $0.6 \%$ at the absolute scale it is difficult to speculate
about its observability just now: its scale is comparable to the widths of the
lines in our plots. However, if we think about it as an effect of the atomic
surrounding on nuclear decay properties, then such a value can be considered as
an impressive one.

It should be noted that the existence of the screening effects is the subject of
the experimental technique employed. For example, the $2p$ decay in $^{19}$Mg 
was
studied in the decay-in-flight experiment Ref.\ \cite{muk07}. In this experiment
the $^{19}$Mg g.s.\ was populated by the neutron knockout from the relativistic
beam of the completely stripped $^{20}$Mg ions. The resulting $^{19}$Mg is also
completely stripped and can hardly pick up any electrons before the decay.
Therefore, in spite of a long lifetime ($T_{1/2}=4$ ps, which is much longer
than typical recombination time), screening in this experiment will have
different character, compared to the case discussed above for $^{45}$Fe.

\subsection{Self-similar solutions}
\label{sec:ss-solutions}


From Figs.\  \ref{fig:mg-traj} and  \ref{fig:fe-traj}, it is possible to see 
that
there exist so called ``stationary points'' in the kinematical
$\{\varepsilon,\cos(\theta_k)\}$ plane in the ``T'' system. For such points, the
classical trajectories in this plane have zero length. For the degenerate
situation $\varepsilon \equiv 1$, the stationary behavior is trivial; this
situation is not very interesting as the phase space for such configurations in
the quantum-mechanical problem tends to zero. However there exist
nondegenerate stationary points,  and which for two-proton decay with a heavy 
core
$\{A_3,Z_3\}$ are found as:
\begin{equation}
\varepsilon = \frac{(1+A_3/2)(Z_3/A_3)^{2/3}}{2A_3+(Z_3/A_3)^{2/3}}\,,\quad
\cos(\theta_k)=0 \,.
\label{eq:stationary}
\end{equation}
These stationary points are defined by the condition that the force acting on
each particle is always directed exactly along the line connecting that particle
with the center of mass of the whole three-body system. Such stationary points
should exist for any two-body potential with the same power dependence on radius
$V(r) \sim r^n$ for each pair of the particles. The values of $\varepsilon$
equal to 0.497, 0.382, 0.379 are found for $^{6}$Be, $^{19}$Mg, and
$^{45}$Fe, respectively, by Eq.\ (\ref{eq:stationary}) as well as by a
direct calculations using
Eq.\ (\ref{eq:newton}). It is clear that the solution, which is a stationary one
in the $\{\varepsilon,\cos(\theta_k)\}$ plane, is an analogue of the Lagrange
solution in celestial mechanics (with reservation  that we are dealing here with
repulsive $1/r$ potentials).

The multi-cluster decays of nuclear systems has been qualitatively studied in
Ref.~\cite{kar04}. In this work a quasiclassical approach was used, based on
the classical self-similar solutions of the few-body Coulomb problem. The
stationary point discussed above represents such a self-similar solution in our
specific case. It was concluded in Ref.\ \cite{kar04} that ``Three-cluster
configuration  asymptotically  approaches  to  an  expanding  self-similar
triangle whose sides obey the $(M/Z)^{1/3}$ rule.'' This statement is probably
not completely correct. It can be seen from Figs.\ \ref{fig:mg-traj} and
\ref{fig:fe-traj} that there is a trend for classical trajectories to tend
somehow towards the stationary point which corresponds to a self-similar
solution. This trend leads to certain systematic modifications of the momentum
distributions by the long-range Coulomb interaction. However as we have seen in
this work, the whole picture is more complex. The total distributions occupy
broad regions of the kinematical plane. They are determined mainly by the
internal structure of the three-body system and the decay dynamics under the
barrier, than by the long-range Coulomb interaction outside of the barrier.
Classical trajectories originating on the hypersphere of large radius each 
converge
to its own final position which, in a general case, could have nothing in
common with a stationary point.


\section{Conclusion}


In this work we discuss the extrapolation along the classical trajectories as a
method to improve the momentum distributions for  radioactive $2p$ decay
(true three-body decay). The proposed method provides near perfect description
of the distributions in the test cases of simplified three-body
Hamiltonians. In the case of real three-body Coulomb interactions
considerable quantitative effects on the distributions are observed. In the case
of the lightest $2p$-emitter $^{6}$Be this effect is minor, but in the heavier
$2p$-emitters ($^{19}$Mg and $^{45}$Fe) the improvement is essential for the
precise description of the distributions.

It should be emphasized that some aspects of the momentum distributions for $2p$
decays are sensitive to the long-range three-body Coulomb interaction, while the
others are absolutely insensitive. Namely, the angular distribution in the
Jacobi  ``T'' system and the energy distribution in the Jacobi  ``Y'' system are
considerably modified by the classical extrapolation. Two other inclusive
distributions (the energy distribution in the Jacobi  ``T'' system and the 
angular
distribution in the Jacobi  ``Y'' system) are essentially not influenced by the
classical extrapolation. Therefore the long-range part of the three-body Coulomb
does not practically change the information about the internal structure of the
decaying system which is contained in the latter distributions.

Attention should be paid to the huge range which is required both for the 
extrapolation
range ($\sim 10^5$ fm) and for the starting point of the classical procedure 
($\sim
10^3$ fm) at typical decay conditions. The classical procedure is applicable 
only
for distances above $500-1500$ fm (in $\rho$ variable) for the considered set of
$2p$ emitters (which is actually quite representative). The intermediate
distances from $30-100$ fm (where the protons come from under the Coulomb
barrier) to around 1000 fm should to be treated quantum mechanically to obtain
decent results from the classical extrapolation.

We have shown that the electron screening can have a sizable effect on the
momentum distribution in the $2p$ decay of atomic $^{45}$Fe. So, the $2p$
radioactivity belongs to a rare class of nuclear phenomena, which exist on the
borderline with atomic phenomena. There exist examples of weak radioactive decay
modification induced by atomic electrons (e.g.\ due to the energy conditions
making $\beta^-$ decay possible only into bound electron states  \cite{jun92} or
due to the hyperfine effect \cite{lit07}). We think that the sizeable
sensitivity of the radioactive decay via particle emission due to a modification
of the potential barrier properties in the atomic environment is demonstrated in
our work for the first time.


\section{Acknowledgments}


%

%

%

L.V.G.\ acknowledges the support from Deutsche Forschungsgemeinschaft grant 436
RUS 113/907/0-1, FAIR-Russia Research Center grant, Russian Foundation for Basic
Research grants RFBR 08-02-00892, RFBR 08-02-00089-a, and Russian Ministry of
Industry and Science grant NSh-7235.2010.2.\\




\end{document}